\newif\ifpreprint
\preprinttrue 

\newif\ifSupplementary
\Supplementaryfalse
\ifpreprint
\documentclass{natureprintstyle}
\else
\documentclass{nature}
\fi
\usepackage{subfigure}
\usepackage{graphicx}
\usepackage{pdfpages}
\usepackage{multirow}
\usepackage{amssymb,hyperref}
\def\msun{$M_{\odot}$} 
\def\lsun{$L_{\odot}$} 
\def\rsun{$R_{\odot}$}

\newcommand\ion[2]{#1$\;${\small\rmfamily #2}\relax}%
\def\araa{Annu.~Rev.~Astron.~Astrophys.}
\def\aap{Astron.~Astrophys.}            
\def\aj{Astron.~J.}                     
\def\apj{Astrophys.~J.}                 
\def\apjl{Astrophys.~J.}                
\def\apjs{Astrophys.~J.~Suppl.~Ser.}    
\def\mnras{Mon.~Not.~R.~Astron.~Soc.}   
\def\nat{Nature}                        
\def\pasp{Publ.~Astron.~Soc.~Pacif.}    
\def\pasj{Publ.~Astron.~Soc.~Japan}    
\def\ssr{Space~Science~Reviews}

\def\aapr{Astron.~Astrophys.~Rev}            
\def\aaps{Astron.~Astrophys.~Suppl.}

\usepackage{multibib}
\newcites{maintext}{\mbox{ }}
\newcites{method}{\mbox{ }}
\newcites{supplementary}{\mbox{ }}

\newcommand\arcmin{\mbox{$^\prime$}}%

\title{A luminous X-ray outburst from an intermediate-mass black hole in an off-centre star cluster}

\author{Dacheng Lin$^{1}$,
Jay Strader$^{2}$,
Eleazar R. Carrasco$^{3}$,
Dany Page$^{4}$,
Aaron J. Romanowsky$^{5,6}$,
Jeroen Homan$^{7,8}$,
Jimmy A. Irwin$^{9}$,
Ronald A. Remillard$^{10}$,
Olivier Godet$^{11}$,
Natalie A. Webb$^{11}$,
Holger Baumgardt$^{12}$,
Rudy Wijnands$^{13}$,
Didier Barret$^{11}$,
Pierre-Alain Duc$^{14}$,
Jean P. Brodie$^{6}$,
Stephen D. J. Gwyn$^{15}$
}

\begin{document}

\maketitle

\begin{affiliations}
\item Space Science Center, University of New Hampshire, Durham, NH 03824, USA
\item Center for Data Intensive and Time Domain Astronomy, Department of Physics and Astronomy, Michigan State University, 567 Wilson Road, East Lansing, MI 48824, USA
\item Gemini Observatory/AURA, Southern Operations Center, Casilla 603, La Serena, Chile
\item Instituto de Astronom\'\i a, Universidad Nacional Aut\'onoma de M\'exico, M\'exico C.D.M.X., 04510, M\'exico
\item Department of Physics and Astronomy, San Jos\'{e} State University, One Washington Square, San Jos\'{e}, CA 95192, USA
\item University of California Observatories, 1156 High Street, Santa Cruz, CA 95064, USA
\item Eureka Scientific, Inc., 2452 Delmer Street, Oakland, California 94602, USA
\item SRON, Netherlands Institute for Space Research, Sorbonnelaan 2, 3584 CA Utrecht, The Netherlands
\item Department of Physics and Astronomy, University of Alabama, Box 870324, Tuscaloosa, AL 35487, USA
\item MIT Kavli Institute for Astrophysics and Space Research, MIT, 70 Vassar Street, Cambridge, MA 02139-4307, USA
\item IRAP, Université de Toulouse, CNRS, UPS, CNES, Toulouse, France
\item School of Mathematics and Physics, University of Queensland, St Lucia, Queensland 4068, Australia
\item Anton Pannekoek Institute for Astronomy, University of Amsterdam, Postbus 94249, 1090 GE Amsterdam, The Netherlands
\item Universit\'{e} de Strasbourg, CNRS, Observatoire astronomique de Strasbourg, UMR 7550, F-67000 Strasbourg, France
\item Canadian Astronomy Data Centre, Herzberg Institute of Astrophysics, 5071 West Saanich Road, Victoria, British Columbia, V9E 2E7, Canada

\end{affiliations}

\begin{abstract}
A unique signature for the presence of massive black holes in very
  dense stellar regions is occasional giant-amplitude outbursts of
  multiwavelength radiation from tidal disruption and subsequent
  accretion of stars that make a close approach to the black
  holes\cite{re1988}. Previous strong tidal disruption event (TDE)
  candidates were all associated with the centers of largely isolated
  galaxies\cite{koba1999,gechre2012,zabema2013,mikami2015,liguko2017}. Here
  we report the discovery of a luminous X-ray outburst from a massive
  star cluster at a projected distance of 12.5~kpc from the center of
  a large lenticular galaxy. The luminosity peaked at
  $\sim$$10^{43}$~erg~s$^{-1}$ and decayed systematically over 10
  years, approximately following a trend that supports the
  identification of the event as a TDE. The X-ray spectra were all
  very soft, with emission confined to be $\lesssim$$3.0$~keV, and
  could be described with a standard thermal disk. The disk cooled
  significantly as the luminosity decreased, a key thermal-state
  signature often observed in accreting stellar-mass black holes. This
  thermal-state signature, coupled with very high luminosities,
  ultrasoft X-ray spectra and the characteristic power-law evolution
  of the light curve, provides strong evidence that the source
  contains an intermediate-mass black hole (IMBH) with a mass of a few
  ten thousand solar mass. This event demonstrates that one of the
  most effective means to detect IMBHs is through X-ray flares from
  TDEs in star clusters.

\end{abstract}

We discovered the X-ray source 3XMM~J215022.4$-$055108 (referred to as
J2150$-$0551 hereafter) in our project of searching for TDEs from
the \emph{XMM-Newton} X-ray source catalog. The source was
serendipitously detected in 2006--2009 in two {\it XMM-Newton}
observations and one {\it Chandra} observation of a field in the
second Canadian Network for Observational Cosmology Field Galaxy
Redshift Survey\cite{cowifi2012}. It was still detected, but with much
lower X-ray fluxes, in our follow-up \emph{Swift} observation in 2014
and \emph{Chandra} observation in 2016. Figure~1 shows
the \emph{Hubble Space Telescope} (\emph{HST}) ACS F775W image around
the field of J2150$-$0551 taken in 2003. The source lies at an angular
offset of 11.6 arcsec from the center of the barred lenticular galaxy
6dFGS gJ215022.2-055059 (referred to as Gal1 hereafter) and is
spatially coincident with a faint optical object. Gal1 is at a
redshift of $z=0.055$ or a luminosity distance of $D_L=247$ Mpc (for
$H_0=70$~km~s$^{-1}$~Mpc$^{-1}$, $\Omega_\mathrm{M}=0.3$,
$\Omega_{\Lambda} = 0.7$). The chance probability for J2150$-$0551 to
be within 11.6 arcsec from the center of a bright galaxy like Gal1 is
very small (0.01\%, see \emph{SI}), strongly supporting the
association of J2150$-$0551 with Gal1.

The fits to the X-ray spectra with a standard thermal thin disk are
shown in the lower panels in Figure~2 (see also
Table~\ref{tbl:spfit}), and the inferred long-term evolution of the
bolometric disk luminosity is plotted in the upper panel. The
dependence of the bolometric disk luminosity on the apparent maximum
disk temperature is plotted in Figure~3. The most striking feature is
that the disk luminosity approximately scales with the temperature as
$L\propto T^4$ (i.e., a constant inner disk radius, the solid line in
the figure), as often observed in accreting stellar-mass black holes
in the thermal state, during which a standard thermal thin disk
dominates the X-ray emission\cite{remc2006,dogiku2007}. The disk
cooled significantly, with the disk temperature decreasing from 0.28
to 0.14 keV, as the disk luminosity decreased by one order of
magnitude from $1.1\times10^{43}$ to $1.1\times10^{42}$ erg s$^{-1}$
within 10 years. Therefore we have conclusive evidence that most
observations were in the thermal state. The brightest observation (X1
hereafter), taken by \emph{XMM-Newton} in 2006, shows a significant
deviation from the $L\propto T^4$ relation traced out by the other
observations. The X-ray spectrum in this observation seems to have a
disk temperature profile more characteristic of a super-Eddington
accretion state (see \emph{SI}), in which the radiation pressure
is stronger than the gravitational force in the inner disk. Based on
fits with a more physical disk
model \textit{optxagnf}\cite{dodaji2012} to the thermal-state
observations, we infer an IMBH in J2150$-$0551, with a mass between
$5\times10^4$ and $10^5$ \msun, depending on the spin parameter
assumed (see \emph{SI}).

An optical flare was detected between May and November in 2005, before
the X-ray detections (Figure~2), with the source appearing significantly bluer
and brighter --- by $\sim$0.8 mag in the $g\arcmin$ band --- than in
August 2000. Enhanced optical emission was not observed in
the \emph{HST} image in September 2003 (see \emph{SI}). Therefore we
constrain the outburst start time of J2150$-$0551 to be between
September 2003 and May 2005.

The fit to the quiescent photometry of the optical counterpart to
J2150$-$0551 in 2000--2003 with a single stellar population
model\cite{ma2005} implies a star cluster with a stellar mass
$\sim$$10^7$ \msun\ and a bolometric stellar luminosity of
$\sim$$10^7$ \lsun\ (see \emph{SI}). The optical source is spatially
unresolved in the \emph{HST} F775W image, with the half-light radius
estimated to be $\lesssim20$ pc (see \emph{SI}). All these properties
suggest a massive star cluster. It could be a very massive globular
cluster\cite{mihimi2012} or more likely a remnant nucleus of a tidally
stripped dwarf galaxy in a minor merger\cite{drgrhi2003,pfba2013},
given that the galaxy might be in an epoch of frequent minor mergers
--- there seems to be a minor merger of a satellite galaxy with Gal1
near J2150$-$0551 (Figure~1).

The standard TDE theory\cite{re1988,ph1989} predicts the mass
accretion rate to decay with the time $t$ after the stellar disruption
as $t^{-5/3}$. A simple TDE model for the luminosity evolution of
J2150$-$0551 is shown as a solid line in the upper panel in Figure~2,
indicating that the luminosity of the source decayed approximately as
$t^{-5/3}$.  The time when the star was disrupted is inferred to be
around mid October 2003, which is consistent with the constraint on
the outburst start time from the optical variability. Prior to the
observation X1, the mass accretion rate was most likely
super-Eddington, in which case the luminosity would be maintained at
around the Eddington limit owing to the effects of photon trapping and
mass outflows\cite{krpi2012}. Therefore, we assume a constant
luminosity before X1. This Eddington-limited plateau is supported by
the non-detection of the source in an
\emph{XMM-Newton} slew observation in 2004 and also by the low
variability of the optical flux in the flaring phase in 2005 (see
\emph{SI}). Further assuming that the rise was fast and occurred one
month after the disruption\cite{gura2015}, we estimate the total
energy released until the last \textit{Chandra} observation to be
$8.9\times10^{50}$ ergs, and the corresponding mass accreted into the
black hole was $0.061(0.1/\eta)$ \msun, where $\eta$ is the efficiency
in converting rest mass into radiated energy in the sub-Eddington
accretion phase. These values of released energy and accreted mass are
typical of other known TDEs\cite{liname2002,kohasc2004,vaanst2016},
but few TDEs\cite{liguko2017} are known to sustain super-Eddington
accretion rates for long periods, as J2150$-$0551 likely did.

The optical counterpart to J2150$-$0551 is very faint, and we observed
no clear emission lines or absorption features in our recent Gemini
observation. Although we cannot determine its redshift through
spectroscopy, we can securely rule out alternative explanations for
the source based on its unique X-ray spectral properties (see
\emph{SI}). Only a neutron star in a low-mass X-ray binary thermally
cooling right after the crust being heated in a large accretion
outburst could mimic the X-ray spectral evolution observed in
J2150$-$0551. However, this explanation is strongly disfavored,
given the absence of a large accretion outburst in the light curve from the
All-sky Monitor onboard the \emph{RXTE} (see \emph{SI}).

Because J2150$-$0551 was discovered through our systematic search over
the \emph{XMM-Newton} X-ray source catalog, we can estimate the rate
of off-center TDEs like J2150$-$0551 and find it to be
$\sim$10$^{-8}$~Mpc$^{-3}$~yr$^{-1}$ (see \emph{SI}). The TDE rate for
black holes with masses of a few $10^4$ \msun\ in star clusters was
predicted\cite{bamaeb2004,brbakr2011,stme2016} to be between
$10^{-5}$--$10^{-3}$~yr$^{-1}$ per black hole. Then our discovery of
J2150$-$0551 implies a significant number of off-center IMBHs with
masses of a few $10^4$ \msun\ in the local Universe, with a space
density between $\sim$$10^{-5}$--$10^{-3}$~Mpc$^{-3}$.

Our study of the \emph{XMM-Newton} X-ray source catalog also led to
the discovery of a possible TDE associated with the nucleus of an
isolated dwarf galaxy\cite{ligoho2017} that could contain a black hole
with a mass of $10^4$--$10^5$~\msun. The nuclei of dwarf galaxies
probably have similar stellar densities and thus similar TDE rates as
seen in off-center very massive ($\sim$$10^7$ \msun) star
clusters\cite{nokafo2014}. Then our discovery of similar numbers of
centered and off-center TDEs for small black holes with masses of a
few $10^4$ \msun\ could imply similar numbers of centered and
off-center black holes in this mass range\cite{sevami2014}.

A large majority of IMBHs might be electromagnetically invisible if
they preferentially form in dense star clusters, which tend to be
devoid of gas\cite{kibalo2017}. Indeed, a wide variety of searching
strategies have found very few strong IMBH candidates with masses of
$10^2$--$10^5$ \msun. \mbox{ESO~243-49} HLX-1, which also lies in a
lenticular galaxy, was the only one clearly showing very similar X-ray
spectral evolution --- especially the $L\propto T^4$ scaling relation
for the disk --- to that observed in accreting stellar-mass black
holes, except that ESO 243-49 HLX-1 had three orders of magnitude
higher luminosities and much lower disk temperatures as expected for
an IMBH\cite{faweba2009,sefali2011,goplka2012} with a mass of
$\sim$$10^4$~\msun. J2150$-$0551 and ESO 243$-$49 HLX-1 followed a
very similar $L$---$T^4$ scaling relation (Figure~3), implying that
they host black holes with very similar masses (see \emph{SI}). One main
difference is that ESO 243$-$49 HLX-1 showed frequent X-ray outbursts
and was unlikely to be due to a single disruption of a star. Our event
demonstrates that IMBHs off-center from their primary host galaxies
may generate TDEs if they reside in dense star clusters. Because TDEs
of IMBHs are expected to easily reach the maximum luminosity, i.e.,
the Eddington limit\cite{gura2015}, as seen in our event, they provide
a powerful way to detect super-Eddington accreting IMBHs to a large
distance.

\begin{addendum}

\item[Acknowledgments] D.L.  is supported by the National Aeronautics
  and Space Administration through Chandra Award Number GO6-17046X
  issued by the Chandra X-ray Observatory Center, which is operated by
  the Smithsonian Astrophysical Observatory for and on behalf of the
  National Aeronautics Space Administration under contract NAS8-03060,
  and by the National Aeronautics and Space Administration ADAP grant
  NNX17AJ57G.  A.J.R. was supported by National Science Foundation
  grant AST-1515084, and as a Research Corporation for Science
  Advancement Cottrell Scholar. J.S. acknowledges support from NSF
  grant AST-1514763 and a Packard Fellowship. D.P. was partially
  supported by the Consejo Nacional de Ciencia y Tecnología with
  CB-2014-1 Grant No. 240512. NW, OG and DB acknowledge CNES for
  financial support to the XMM-Newton Survey Science Center
  activities. R.W. acknowledges support from the Netherlands
  Organisation for Scientific Research through a Top Grant, module 1.
  J. B.  acknowledges support from National Science Foundation grant
  AST 1518294.  We thank the former \textit{Swift} PI Neil Gehrels for
  approving our ToO request to make an observation of J2150$-$0551. We
  thank Zachary Jennings for assistance with the Suprime-Cam
  data. Based on observations obtained from XMM-Newton, Chandra,
  Swift, HST, CFHT, Gemini, SOAR, and Subaru.

\item[Author Contributions]

  D.L.  wrote the main manuscript and led the data analysis.
  E.C. helped reduce the GMOS spectra and pre-imaging. D.P. performed the
  MCMC simulations for NSCool. J.S. obtained the SOAR $U$-band image
  and fitted the \emph{HST} image with ISHAPE. A.R. obtained the Subaru
  $g\arcmin$-band image. S.G. stacked the CFHT images. All authors
  discussed the results and commented on the manuscript.

\item[Competing Interests] 
The authors declare that they have no competing financial interests.

\item[Author Information] 
Correspondence and requests for materials should be addressed to
D.L. (dacheng.lin@unh.edu). 
\end{addendum}

\clearpage

\begin{table*}
\centering
\caption{\textbf{Fitting results of the high-quality X-ray spectra
    from J2150$-$0551 with an absorbed \textit{diskbb} model.} X1 and X2 are
  \emph{XMM-Newton} observations, while C1 and C2 are \emph{Chandra}
  observations (refer to Supplementary Table 1). All errors are at the 90\% confidence level. Galactic
  absorption was included and fixed at $N_\mathrm{H,
    Gal}=2.6\times10^{20}$ cm$^{-2}$. $N_\mathrm{H,i}$ is the
  absorption intrinsic to the X-ray source at redshift 0.055. It was
  tied together in the simultaneous fit to all spectra and then fixed
  at the best-fitting value when we calculated the uncertainties of other
  spectral parameters. The \textit{diskbb} parameter $kT_\mathrm{disk}$ is the disk apparent maximum
  temperature, and the normalization $N_\mathrm{disk}$ is defined as $((R_\mathrm{disk}/\mathrm{km})/(D/\mathrm{10 kpc}))^2\cos\theta$,
  where $R_\mathrm{disk}$ is the apparent inner disk radius, $D$ is
  the source distance, $\theta$ is the disk inclination angle.
  $\chi^2_\nu$ is the reduced $\chi^2$ value, and $\nu$ is the degrees
  of freedom. $L_{\rm
    abs}$ is the source rest-frame 0.32--10.6 keV (i.e., observer-frame 0.3--10 keV) luminosity, corrected for the Galactic absorption but not intrinsic absorption,
  and $L_{\rm unabs}$ is the source rest-frame 0.32--10.6 keV luminosity,
  corrected for both Galactic and intrinsic
  absorption. $L_\mathrm{disk}$ is the unabsorbed bolometric disk
  luminosity. We note that in the fits to X1, C1 and X2 we added a
  power-law component of photon index 1.8, in order to account for
  possible contamination from the nuclear source of Gal1; this component
  was found to be very weak and was not included in the luminosity
  calculation (see \emph{SI}). All luminosities assume a disk
  inclination of $60^\circ$.}
\label{tbl:spfit}
\bigskip
\scriptsize
\sffamily
\begin{tabular}{rcccccccc}
\hline
\hline
&X1 & C1 & X2 & C2\\
Date & 2006-05-05 & 2006-09-28 & 2009-06-07 & 2016-09-14\\
\hline
$N_\mathrm{H,i}$ (10$^{20}$ cm$^{-2}$) & \multicolumn{4}{c}{$0.6_{-0.6}^{+1.0}$}\\
$kT_\mathrm{disk}$ (keV) & $0.267\pm0.005$  & $0.279\pm0.009$  & $0.236\pm0.006$ &$0.141\pm0.012$ \\
$N_\mathrm{disk}$ &$15.0\pm1.4$  &$8.6\pm1.1$  &$6.5\pm0.7$ & $19.0^{+14.8}_{-8.9}$\\
$\chi^2_\nu(\nu)$&$1.01(326)$&$1.28(101)$&$0.96(360)$&$1.70(7)$ \\
$L_\mathrm{abs}$ (10$^{42}$ erg~s$^{-1}$)  & $ 6.98^{+ 0.22}_{-0.21}$  & $ 4.87^{+ 0.16}_{-0.15}$   & $ 1.68^{+ 0.08}_{-0.08}$ &$ 0.40^{+ 0.08}_{-0.08}$\\
$L_\mathrm{unabs}$ (10$^{42}$ erg~s$^{-1}$) & $ 7.21^{+ 0.23}_{-0.22}$ & $ 5.03^{+ 0.16}_{-0.16}$  & $ 1.74^{+ 0.08}_{-0.08}$ & $ 0.42^{+ 0.09}_{-0.08}$ \\
$L_\mathrm{disk}$ (10$^{42}$ erg~s$^{-1}$) & $10.63^{+ 0.35}_{-0.34}$  & $ 7.25^{+ 0.30}_{-0.28}$  & $ 2.76^{+ 0.14}_{-0.13}$ & $ 1.05^{+ 0.32}_{-0.28}$\\
\hline
\hline
\end{tabular}
\end{table*}

\clearpage

\begin{figure*}
\begin{center}
\includegraphics[width=3.5in]{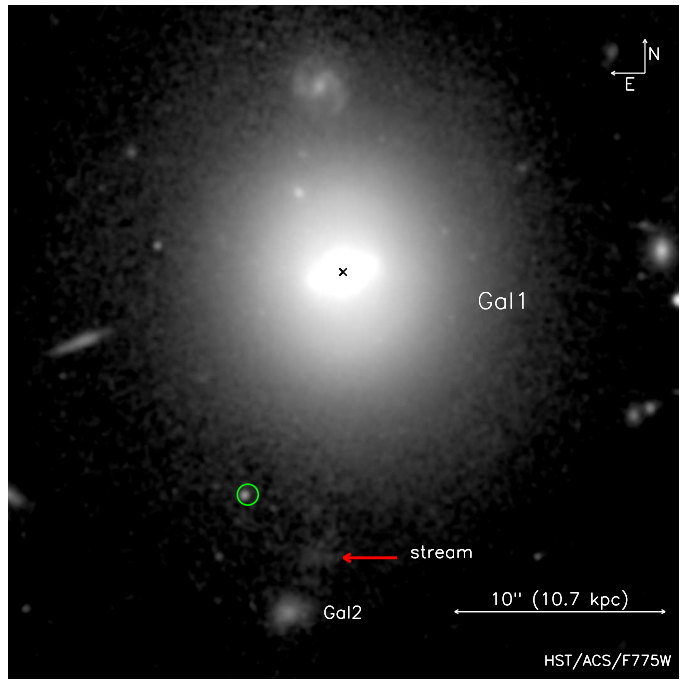}
\end{center}
\vskip -0.2in
\caption{
  \textbf{The {\it HST}/ACS F775W imaging around the field of
    J2150$-$0551.} The image was smoothed with a 2-D gaussian function
  of $\sigma=0.1$ arcsec. J2150$-$0551 appears to be in a barred
  lenticular galaxy Gal1.  The X-ray position of J2150$-$0551 from the
  \emph{Chandra} observation C2 is marked with a green circle, whose
  radius, for clarity, is twice as large as the 99.73\% X-ray
  positional error (0.25 arcsec). The source has a faint optical
  counterpart, at an offset of only 0.14 arcsec from the X-ray position.
  The galaxy at the bottom of the image (Gal2) could be a satellite
  galaxy connected with Gal1 through a tidal stream (red arrow),
  indicating that Gal1 is rich in minor mergers. \label{fig:hstimg}}
\end{figure*}

\clearpage

\begin{figure*}
\begin{center}
\includegraphics[width=5.8in]{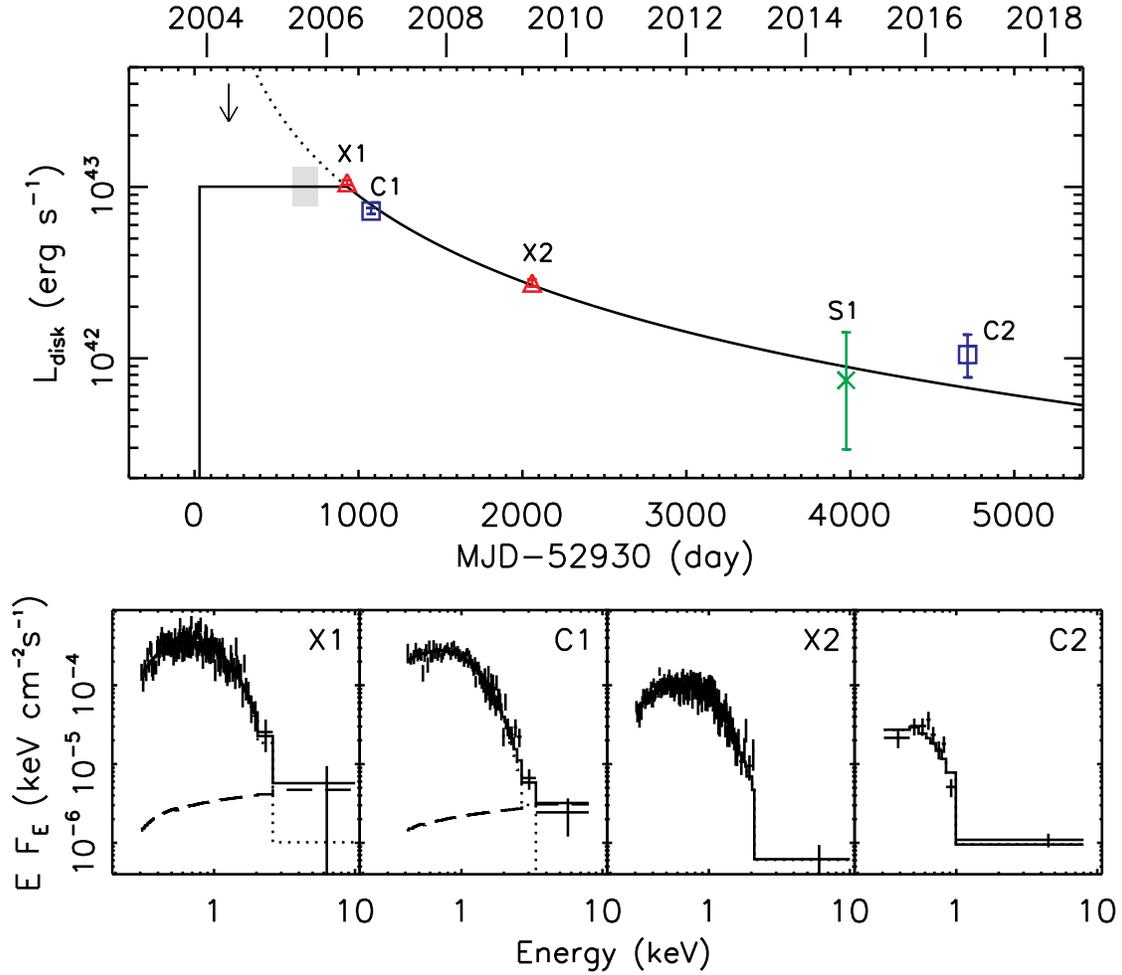}
\end{center}
\vskip -0.2in
\caption{\textbf{The long-term luminosity and spectral evolution of
    J2150$-$0551.} Upper panel: the bolometric disk luminosity curve.
  The \emph{Chandra}, \emph{XMM-Newton} and \emph{Swift} pointed
  observations are shown as blue squares, red triangles and a green
  cross, respectively, with 90\% error bars, and the arrow shows the
  $3\sigma$ upper limit from the \emph{XMM-Newton} slew observation on
  14 May 2004. The gray shaded region marks the time interval when the
  optical flare was detected in 2005.  The solid line is a simple
  luminosity evolution model, with the luminosity following a
  $(t-t_{\rm D})^{-5/3}$ decline after X1 and the disruption time
  $t_{\rm D}$ on 18 October 2003. The rise was assumed to be fast,
  occurring one month after disruption, and the luminosity before X1
  was assumed to be constant, given that the source was most likely in
  the super-Eddington accretion phase during that time. The dotted
  line neglects this Eddington-limited effect and predicts a
  luminosity a factor of 3 higher than the $3\sigma$ upper limit
  implied by the \emph{XMM-Newton} slew observation. Lower panels: the
  standard thermal disk fits to the X-ray spectra in different
  observations. A power-law component (dashed line) was added in the
  fits to X1, C1 and X2 (inferred to be zero in this observation) to
  account for possible contamination of the nuclear source of Gal1 in
  these observations, which had relatively large point spread
  functions. For clarity, the spectra are rebinned to be above
  2$\sigma$ in each bin in the plot, and for the \emph{XMM-Newton}
  observations, we only show the pn spectra. We did not fit the S1
  spectrum due to its low statistics, and its luminosity was estimated
  based on the spectral fit to C2. \label{fig:lumlcsp}}
\end{figure*}
\clearpage

\begin{figure*}
\begin{center}
\includegraphics{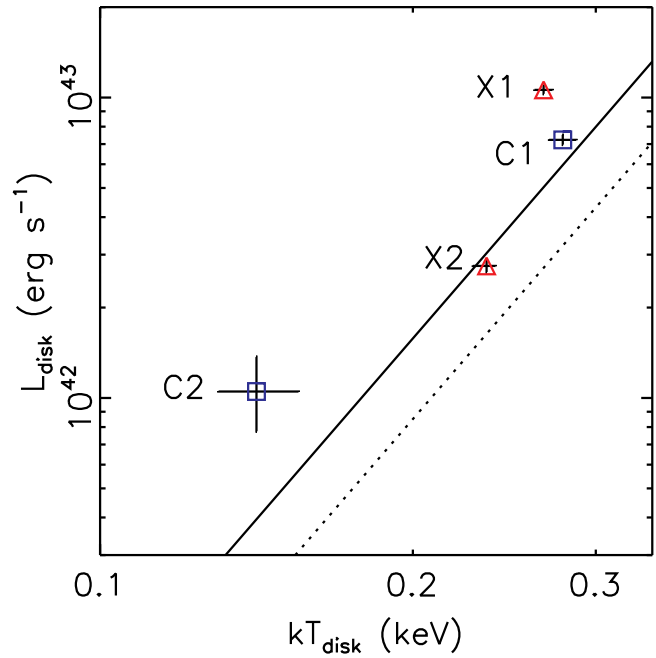}
\end{center}
\caption{\textbf{The disk luminosity versus temperature derived from the
    fits to the X-ray spectra of J2150$-$0551 with the standard
    thermal disk model.} Error bars are at the 90\% confidence level. The solid line plots the $L\propto T^4$
    relation with the inner disk radius being the mean value of those
    in C1, X2, and C2, weighted by the errors. These observations seem
    to follow the $L\propto T^4$ evolution (within $2.2\sigma$), as
    would be expected if they were in the thermal state. X1 deviated
    significantly ($9\sigma$) from this evolution and can be explained
    if it was in the super-Eddington accretion state. The nearby
    dashed line plots the $L\propto T^4$ relation\cite{sefali2011}
    traced out by ESO 243$-$49 HLX-1, suggesting that J2150$-$0551 and
    ESO 243$-$49 HLX-1 contain black holes of similar
    masses. \label{fig:mcdlumkt}}

\end{figure*}

\clearpage

\clearpage

\Supplementarytrue

\section*{Method}

\noindent \textbf{X-ray Observations and Data Analysis}

J2150$-$0551 was serendipitously detected in two \emph{XMM-Newton}
observations (hereafter X1 and X2) and one \emph{Chandra} observation
(hereafter C1, Supplementary Table~\ref{tbl:obslog}) of a field in the
second Canadian Network for Observational Cosmology Field Galaxy
Redshift Survey\cite{cowifi2012}. X1 was taken in May 2006, and X2 in
June 2009. C1 was taken in September 2006. J2150$-$0551 was at
off-axis angles of $\sim$6\arcmin--8\arcmin\ in these observations. We
also obtained three X-ray follow-up observations, two by \emph{Swift}
in September 2014 (hereafter S1 for their combination, Supplementary
Table~\ref{tbl:obslog}) and one by \emph{Chandra} in September 2016.

We used SAS 16.0.0 and the calibration files of March 2017 to analyze
the two \emph{XMM-Newton} observations. The source was in the field of
view (FOV) of all the three European Photon Imaging Cameras (i.e., pn,
MOS1, and MOS2)\cite{jalual2001,stbrde2001,tuabar2001} in the imaging
mode in both observations. Strong background flares were
seen in all cameras in both X1 and X2, and we excluded data in these
intervals following the SAS thread for the filtering against high
backgrounds. The exposure times of the clean data finally used are
listed in Supplementary Table~\ref{tbl:obslog}. We extracted the
source spectra from all cameras using circular regions,
whose radii are given in Supplementary Table~\ref{tbl:obslog}. The
background spectra were extracted from large circular regions with
radii in the range of 50--100 arcsec near the source. We
adopted the event selection criteria used in the
pipeline\cite{wascfy2009}. We used the SAS task \textit{rmfgen} to
create the response matrix files and \textit{arfgen} to create the
point-source aperture corrected auxiliary response files. We note that
there were bright columns through the source extraction region in pn
in X1. These columns were excluded after the standard good event
selection, resulting in a significant reduction on the source counts
from pn. In both MOS1 and MOS2 cameras, a small fraction of the source counts
were lost because the source was close to CCD gaps. We also performed
the variability test by applying the SAS tool \emph{ekstest} to the
light curves of both observations extracted with the tool
\emph{epiclccorr}.

We note that there was no obvious pile-up for J2150$-$0551 in the
\emph{XMM-Newton} data because the source was not very bright. This is
supported by the consistent fits obtained with the spectra from
different cameras
(see Supplementary Figure~\ref{fig:srcspec_disk_delchi}). Pile-up, if
present, would affect the MOS1 and MOS2 spectra much more seriously than the pn
spectra. In addition, we also created spectra using only
single-pattern events, which would significantly reduce any possible
pile-up, but we obtained consistent fitting results, compared with the
fits to the spectra created above using all standard good events.

The C1 observation used the AXAF CCD Imaging Spectrometer
(ACIS)\cite{bapiba1998}, with the aimpoint and J2150$-$0551 falling in
the front-illuminated chips I3 and I0, respectively. Our follow-up
observation C2 also used the ACIS, but with the aimpoint falling in
the back-illuminated chip S3 (J2150$-$0551 was offset from the aim
point by only 9 arcsec).  We reprocessed the data to apply the latest
calibration (CALDB 4.7.3) using the script \texttt{chandra\_repro} in
the \emph{Chandra} Interactive Analysis of Observations (CIAO) package
(version 4.9). We used the CIAO task \texttt{specextract} to extract
the spectra of J2150$-$0551 and create the corresponding response
files. We extracted the source spectra using circular regions of radii
7.7 and 1.6 arcsec, enclosing 90\% and 95\% of the point spread
function (PSF) at 1 keV, for C1 and C2, respectively. The background
spectra were extracted from circular regions of radius 40 arcsec
near the source in both observations. We point out that pile-up was
negligible in C1 because J2150$-$0551 was not very bright and was at a
very large off-axis angle of 8 arcmin, resulting in a very large
PSF. Pile-up was not an issue in C2 either, because J2150$-$0551 was
very faint in this observation.

There was clearly a faint source detected in C2 at the nucleus of Gal1
(hereafter Source 1). We extracted the spectra for this source from
this observation as well, in the same way as we did for
J2150$-$0551. We note that Source 1 was not clearly seen in
observations X1, X2, and C1, because Source 1 was much fainter than
J2150$-$0551 and they were hardly resolved from each other due to their
large PSFs in these observations. When we fitted the spectra
  of J2150$-$0551 from X1, X2 and C1 in the \emph{SI}, we had an extra
  spectral component to account for the contamination from Source 1
  and found this nuclear source to be very faint, as observed in C2.

The best position of J2150$-$0551 can be obtained from the on-axis
\textit{Chandra} observation C2, which had the sub-arcsec resolution
near the aimpoint.  We carried out absolute astrometric correction and
determined the statistical and systematic positional errors in the
same way as we did in our previous studies\cite{licawe2016,
  liguko2017}. In brief, we first carried out the X-ray source
detection using the CIAO tool \texttt{wavdetect}\cite{frkaro2002}, and
then performed absolute astrometric correction using 25 matches of
X-ray sources and optical sources from the CFH12K
mosaic\cite{culust2000} $R$-band image (see below).

In the two \emph{Swift} observations in 2014, the X-ray telescope
(XRT)\cite{buhino2005} was operated in Photon Counting mode. We
analyzed the data with FTOOLS 6.20 and the calibration files of May 2017. The X-ray data were reprocessed with the task
\textit{xrtpipeline} (version 0.13.3) to update the calibration. We
extracted the source and background spectra using circular regions of radii
20 and 100 arcsec, respectively.\\

\noindent \textbf{Multi-wavelength Photometry Observations}

In both \emph{XMM-Newton} observations X1 and X2, J2150$-$0551 was
also covered in the images of the Optical Monitor
(OM)\cite{mabrmu2001}, which used only the \textit{UVM2} (2310 \AA)
filter. The source was not detected in these images, whose $3\sigma$
detection limit was $\sim$22.0 AB mag. In both \emph{Swift}
observations in September 2014, the UV-Optical Telescope
(UVOT)\cite{rokema2005} images used the $UVW2$ (2120 \AA) filter.
J2150$-$0551 was not detected either in these images, whose $3\sigma$
detection limit was 24.3 AB mag.

J2150$-$0551 happened to be covered in an image from the Advanced
Camera for Surveys (ACS) Wide Field Camera (WFC) on the \textit{Hubble
Space Telescope} (\textit{HST}) on 13 September 2003 (Supplementary
Table~\ref{tbl:obslogopt}). As will be shown, it clearly revealed a
very faint counterpart to J2150$-$0551. The image was composed of four
510-s exposures, but J2150$-$0551 was covered by only two of them because
it happened to be in the CCD gap (in the north-south direction) in the
other two
exposures. We measured the photometry using a circular source region
of radius 0.5 arcsec and an annular background region of inner radius
0.6 arcsec and outer radius 0.8 arcsec. The photometry was then corrected for the PSF loss\cite{sijebe2005}.

There are also many optical/IR observations in the field of J2150$-$0551
by the CFH12K mosaic\cite{culust2000}, the MegaPrime/MegaCam\cite{bochab2003} and the WIRCam\cite{pustdo2004} on the
Canada-France-Hawaii Telescope (CFHT, Supplementary Table~\ref{tbl:obslogopt}). We
used the calibrated images obtained in the literature\cite{bamcwi2009}. They were
calibrated by comparing with the SDSS photometry. These ground-based
images have much larger PSFs than the \emph{HST} image. Therefore,
there is significant non-uniform stellar emission of the host galaxy
at the position of J2150$-$0551, making the calculation of the
photometry of our source in these images non-trivial. We carried out
detailed fits to the profile of the host galaxy plus the counterpart
to J2150$-$0551 in these images using GALFIT\cite{pehoim2010} in order
to calculate the photometry of J2150$-$0551.  Multiple S\'{e}rsic
functions convolved with the PSF were used to fit the galaxy
profile, while a PSF model was used to fit the counterpart to
J2150$-$0551, which showed no evidence of being resolved in the
\emph{HST} image. Because the main goal was to obtain the photometry of
J2150$-$0551, we tried to use an adequate number of S\'{e}rsic functions
to ensure reasonable fits to the host galaxy emission. The exact
numbers of S\'{e}rsic functions used varied with images and were in
the range between five and nine.

To measure the current optical emission level, we obtained a SOAR
$U$-band image and a Subaru $g$-band image of our source in 2014
(Supplementary Table~\ref{tbl:obslogopt}). We also obtained a Gemini
$r_\mathrm{G0326}$-band image, which was used to design the mask for the
Gemini Multi-Object Spectrograph (GMOS)\cite{hojoal2004} observation
of J2150$-$0551 (program id {\sc GS-2016A-Q-20}). All these images
were reduced and calibrated using standard procedures. The above GALFIT
photometry method was also applied to these images.\\

\noindent \textbf{The GMOS Observation}

A special mask was designed for the GMOS observation of J2150$-$0551,
Gal1, and Gal2 (see Supplementary Figure~\ref{fig:geminiimg}). To take
full advantage of the multi-object spectroscopy capabilities of
the instrument, other galaxies were also included in the mask design,
but the analysis and scientific use of them will be presented in a
separate paper. The GMOS spectroscopic data were obtained during the
nights of 27 September 2016 and 4 November 2016, in dark conditions,
with a good sky transparency and a median seeing between 0.45
and 0.55 arcsec. All spectra were acquired with the R400 grating
centered at 5500~\AA.  A total exposure time of 3 hours
($6 \times 1800$ seconds) was obtained for the mask. Small offsets of
$\sim$50 \AA\ between exposures toward the blue and red were applied
to avoid the loss of any important lines present in the spectra.  All
objects in the mask were observed with 1 arcsec slit widths, except
Gal2, which has a larger size and thus was observed with a 1.5 arcsec
slit width.

The spectroscopic data were reduced with the {\sc Gemini} package
version 1.3.1 inside {\sc IRAF}. The science spectra, comparison lamps
and spectroscopic flats were bias/overscan subtracted, trimmed and
flat fielded using the spectroscopic flats. The 2-D spectra were then
wavelength calibrated, corrected by S-shape distortions and
sky-subtracted. Because our primary interest resided in the galaxies
Gal1 and Gal2 and the optical counterpart to J2150$-$0551, only the
spectra of these three objects were extracted to a one-dimensional
format. For Gal1 and Gal2, we used a fixed aperture of 1.4 arcsec. For
the optical counterpart to J2150$-$0551, a fixed aperture of 0.8 arcsec
was used, with the background taken to be the mean of two background
regions that were on different sides of the source and had the same
size as the source aperture, in order to account for the effect of the
Gal1 stellar emission near J2150$-$0551. The residual values in the
wavelength solution for $\sim$25 points using a 5th-order Chebyshev
polynomial typically yielded \textit{rms} values of
$\sim$0.15~\AA. The final spectral resolution (FWHM, measured from the
sky lines at 6300 \AA) is $\sim$7.1 \AA\ for Gal1 and the optical
counterpart to J2150$-$0551 (1 arcsec slit width) and $\sim$8.2 \AA\
for Gal2 (1.5 arcsec slit width). The wavelength coverage was
$\sim$4200--8000~\AA, depending on the position of the slit in the
GMOS FOV. Finally, the 1D spectra were calibrated in flux using the
observation of the spectrophotometric standard star LTT~1788. Because
this standard star was observed on a different night (2 February 2016)
and under different observing conditions, the science spectra were
only relatively calibrated in flux.

We fitted the GMOS spectra with Penalized Pixel Fitting ({\tt pPXF})
software\cite{caem2004}, which extracts the stellar kinematics or
stellar population from spectra of galaxies with the penalized
pixel-fitting method. We used multi-component models comprised of
single-population synthetic spectra\cite{vasafa2010}, spanning a grid
of 48 ages between 0.06 to 14 Gyr and 7 metallicities [M/H]=\{$-2.32$,
$-1.71$, $-1.31$, $-0.71$, $-0.40$, $0.00$, $+0.22$\}. For the
spectrum of the nuclear region of Gal1, we added several Gaussian
lines to model the possible presence of emission lines. This spectrum
is of high quality, and we incorporated a multiplicative polynomial of
10 degrees to account for possible calibration uncertainties. The
spectra of Gal2 and J2150$-$0551 are noisy, and we did not incorporate
a multiplicative polynomial in the fits. All spectra were corrected for
the Galactic dust reddening\cite{scfida1998} of
$\mathrm{E(B-V)}_\mathrm{G}=0.03$
mag before the fits. \\

\noindent \textbf{Numerical modeling of a cooling neutron star scenario for J2150$-$0551}

As will shown in the \emph{SI}, the very soft X-ray spectra of J2150$-$0551
approximately resemble those of thermally cooling neutron stars (NSs)
in low-mass X-ray binaries (LMXBs) often detected after large
outbursts due to accretion\cite{hofrwi2014, deoore2017}. Cooling
curves have been observed in about ten NS LMXBs\cite{widepa2017}, and
they were generally explained as the cooling of the NS crust after
being heated by non-equilibrium nuclear reactions during large
accretion outbursts\cite{hazd1990,brbiru1998}. We need to estimate the
mass accretion rate $\dot M$ needed to produce the hypothetically
observed cooling curve of J2150$-$0551 and compare this $\dot M$ with
the detection limit from the All-Sky Monitor
(ASM)\cite{lebrcu1996} onboard the \emph{RXTE}, in order to check the
feasibility of the cooling NS explanation for J2150$-$0551. We
estimated $\dot M$ through extensive Markov chain Monte Carlo (MCMC)
simulations with the code
\texttt{NSCool}\cite{palapr2004,pare2013,pa2016}, using an MCMC driver
inspired from the \texttt{emcee} code\cite{fohola2013}. We used 200
``walkers'' exploring simultaneously the parameter space. The
parameters of the MCMC runs covered a wide range of possibilities
about the structure and microphysics of the NS and the characteristics
of the accretion outburst responsible for the heating of the NS prior
to the four observations X1, C1, X2, and C2. We modeled the accretion
phase as a step function in time with a constant $\dot M$, with the
start time constrained to be between 14 May 2004 and 3 June 2005 and
the end time to be between 7 November 2005 and 5 May 2006 (see
\emph{SI}). We explored a wide range of possible values of the NS mass
($M_\mathrm{NS}$=1.0--2.2 \msun) and radius ($R_\mathrm{NS}$=8--16
km). The cooling curve and the source distance inferred from the
spectral fits depend on the NS mass and radius assumed. We first
obtained the cooling curves and the source distances for a grid of 13
NS masses (step 0.1 \msun) and 9 radii (step 1 km) from the spectral
fits and then interpolated them to obtain the cooling curve and source
distance for any NS mass and radius of each MCMC run. In order to
convert $\dot M$ from the MCMC to the ASM count rate, we assumed a
radiative efficiency of $1-\sqrt{1-2GM_\mathrm{NS}/R_\mathrm{NS}c^2}$,
where $c$ is the speed of light and $G$ is the gravitational constant,
and adopted a bolometric correction factor of five from the ASM band
pass (1.5--12 keV) to 0.2--100 keV, based on typical low-state spectra
of NS LMXBs\cite{ardewi2013}.

Shallow heating of the crust released at low densities
($\sim$$10^8$--$10^{10}$ g cm$^{-3}$), beyond the deep crustal
heating, is often needed in explaining the cooling curves of many
objects\cite{brcu2009,pare2013,decubr2015}. Therefore we also
considered this possibility and added a shallow energy source, which
was parametrized in terms of the amount of energy released per
accreted nucleon, $Q_\mathrm{sh}$. Other parameters included in the
MCMC were the amount of light elements in the stellar envelope, the
impurity parameter that controled electron scattering for the thermal
conductivity, and the extent of neutron superfluidity in the inner
crust that strongly affected the specific heat. \\

\noindent \textbf{Data availability statement}
The data that support the plots within this paper and other findings of this study are available from the
corresponding author upon reasonable request.

\clearpage
\ifpreprint
\else
\setcounter{page}{1}
\fi
\setcounter{figure}{0}
\setcounter{table}{0}
\renewcommand{\thefigure}{\arabic{figure}}
\renewcommand{\thetable}{\arabic{table}}

\Supplementarytrue

\section*{Supplementary Information}

\mbox{ }

\vskip 0.05in 
\noindent

\noindent \textbf{The X-ray Source Position and the Optical Counterpart}

Figure~\ref{fig:hstimg} shows the \textit{HST}/ACS image of the field
around J2150$-$0551, which is marked with a green circle of radius
twice as large as its 99.73\% X-ray positional error (0.25 arcsec) that
we obtained from C2. J2150$-$0551 appears to be on the outskirts of
Gal1, a barred lenticular galaxy at $z=0.05526\pm0.00003$ (from the
pPXF fit below, $D_{\rm L}=247$ Mpc, and the 6dF Galaxy Survey gave a
consistent redshift\cite{joresa2009}). J2150$-$0551 is at an angular
offset of 11.6 arcsec, corresponding to 12.5 kpc, from the nucleus of
Gal1. J2150$-$0551 is at an offset of only 0.14 arcsec (less than the
99.73\% positional error) from a faint optical source at
R.A.=21:50:22.498 and Decl.=$-5$:51:09.093. Further considering the
correlated optical and X-ray variability, as will be demonstrated
below, we can securely identify this optical source as the counterpart
to J2150$-$0551. The number density of galaxies that are as bright as
or brighter than Gal1 in the Two Micron All Sky
Survey\cite{skcust2006} $K_\mathrm{s}$ band (12.3 mag) near
J2150$-$0551 is 3.9 deg$^{-2}$. Then the chance probability for
J2150$-$0551 to be within 11.6 arcsec from the nucleus of Gal1 is very
low, only 0.01\%. Therefore, Gal1 is most likely the host of
J2150$-$0551. Figure~\ref{fig:hstimg} shows that near J2150$-$0551
there appears to be a smaller galaxy (marked as ``Gal2''), which could
be connected with Gal1 with a stream and thus could be a satellite
galaxy of Gal1. The possible stream is also seen in various
ground-based images, such as the Gemini $r_\mathrm{G0326}$ image shown
in Supplementary Figure~\ref{fig:geminiimg}. Therefore Gal1 could be
rich in minor mergers.

To test whether the counterpart to J2150$-$0551 was resolved in the
\emph{HST} image, we used {\tt ishape}\cite{la1999}. We created an
empirical PSF using several stars in the field and
fitted a range of models, including both King and S\'{e}rsic functions. All
fits were performed on the two individual flat-fielded images in which
the source appears, rather than on the final drizzled image. For no
model tested was there a significant improvement in the reduced
$\chi^{2}$ of the fit over a point source. While deeper data are
desirable, we found no compelling evidence that the source was resolved
in the available images. If the object was extended, its effective
radius would be $\lesssim$0.02 arcsec, or 20 pc.

C1 also detected a faint X-ray source (Source 1) at an offset of only
0.06 arcsec from the optical center of Gal1. The offset is much smaller
than the 99.73\% X-ray positional error of Source 1 (0.39 arcsec or 0.42
kpc), supporting the nuclear origin
for the source. \\

\noindent \textbf{X-ray Spectral Modeling}

The \emph{XMM-Newton} and \emph{Chandra} spectra (X1, C1, X2, and C2)
of J2150$-$0551 are of good quality, and we fitted them in detail. The
spectra were rebinned to have a minimum of 15 counts per bin in order
to adopt the $\chi^2$ statistic in the fits in
XSPEC\cite{ar1996}. Because J2150$-$0551 most likely resides in the
galaxy Gal1 at $z=0.055$, we first explored this scenario by applying
this redshift to all the spectral models that we tested with the
convolution model \textit{zashift} in XSPEC. Galactic
absorption\cite{kabuha2005} of $N_\mathrm{H}=2.64\times10^{20}$
cm$^{-2}$ was included using the \textit{tbabs} model, while
absorption intrinsic to the source was modeled with
\textit{ztbabs}. The Wilm abundance table\cite{wialmc2000} was
adopted. Because the X1, C1, and X2 spectra were obtained from
relatively large source regions due to large PSFs in these
observations, they might be very weakly contaminated by the nearby
nuclear emission Source 1. The spectrum of Source 1 in C2 can be
fitted with an absorbed power law (PL) of Galactic absorption
(intrinsic absorption is consistent with zero) and
$\Gamma_\mathrm{PL}=1.8\pm0.5$ (the source rest-frame 0.32--10.6 keV
unabsorbed luminosity was $6\times10^{40}$ erg s$^{-1}$). Therefore,
in our fits to the X1, C1, and X2 spectra, we had an extra absorbed PL
component with $\Gamma_\mathrm{PL}$ fixed at $1.8$ and $N_\mathrm{H}$
fixed at $2.64\times10^{20}$ cm$^{-2}$ to account for possible
contamination from Source 1. The inferred contamination was found to
be very weak and have negligible effects on the fitting results of
J2150$-$0551.

All spectra of J2150$-$0551 were very soft, with little emission above
3 keV. The first spectral model that we tested was an absorbed PL. We
obtained very high photon indices ($\Gamma_\mathrm{PL}\ge4.8$) for all
spectra. The fits showed clear systematic residuals (e.g., reduced
$\chi^2$ value $\chi^2_\nu$=1.26 for degrees of freedom $\nu=325$ for
X1). We next tried to fit the spectra with an absorbed
single-temperature blackbody (\textit{bbodyrad} in XSPEC). However,
the fits were generally bad (e.g., $\chi^2_\nu$=1.45 for $\nu=325$ for
X1 and $\chi^2_\nu$=1.54 for $\nu=100$ for C1), with the systematic
residuals indicating that the spectra are broader than a
single-temperature blackbody. We next tried an absorbed multicolor
disk (\textit{diskbb} in XSPEC) model and obtained acceptable fits
with $\chi^2_\nu$ in the range of 0.9--1.3 for X1, C1, and X2. The
$\chi^2_\nu$ value was higher for C2, but with $\nu=6$, the
probability that the spectrum was drawn from the model is modest (9\%).
All spectra inferred a small and consistent intrinsic column density
$N_\mathrm{H,i}$. In the final fits, we fitted all spectra
simultaneously with $N_\mathrm{H,i}$ tied to be the same. The fit
results are given in Table~\ref{tbl:spfit} and are shown in
Figure~\ref{fig:lumlcsp}. The fit residuals are shown in Supplementary
Figure~\ref{fig:srcspec_disk_delchi}. We note that we fixed
$N_\mathrm{H,i}$ at the best-fitting value when we calculated the
uncertainties of the other parameters. Fixing $N_\mathrm{H,i}$ could
put tighter constraints on the disk parameters and allow us to
determine the significance of the relative changes of the disk properties with the
observations more accurately.

Figure~\ref{fig:mcdlumkt} plots the bolometric disk luminosity
$L_\mathrm{disk}$ versus the the apparent maximum disk temperature
$kT_\mathrm{disk}$ from the \textit{diskbb} fits. Remarkably, the disk
luminosity overall approximately follows the $L\propto T^4$ relation
(constant inner disk radius, the solid line in the figure), as
observed in black hole (BH) X-ray binaries in the thermal
state\cite{remc2006,dogiku2007}. The disk cooled significantly, with
$kT_\mathrm{disk}$ decreasing from $\sim$0.28 keV in X1 and C1 to 0.14
keV in C2 as $L_\mathrm{disk}$ decreased by an order of magnitude (see
also Table~\ref{tbl:spfit}). Especially from C1 to X2 in nearly three
years, the apparent inner disk radius $R_\mathrm{disk}$ decreased only
slightly (by 15\%) at the 2.8$\sigma$ confidence level, but
$kT_\mathrm{disk}$ decreased from 0.28 keV to 0.23 keV at the
7.4$\sigma$ confidence level. $R_\mathrm{disk}$ in C2 seemed to be
larger than that in X2 by 71\%, but only at the $2.5\sigma$ confidence
level, while the cooling of the disk from X2 to C2 was very
significant ($10\sigma$). Therefore, we conclude that the source was
in the thermal state in C1, X2, and C2.

The most significant deviation from the $L\propto T^4$ relation was
observed from X1 to C1. $R_\mathrm{disk}$ decreased by 32\% at the
6$\sigma$ confidence level from X1 to C1 in nearly four months, while
$kT_\mathrm{disk}$ was consistent to within the 2$\sigma$ error.  One
explanation for this deviation is that the inner disk reached the
local Eddington limit in X1, which is brighter than C1 by 50\%. In
such a scenario, the cooling from advected photons and/or
radiation-driven mass outflow in the inner disk is expected to be
significant, resulting in a disk temperature profile with respect to
the radius $r$ to deviate from $T\propto r^{-3/4}$ for a standard
thin
disk\cite{ka1980,abczla1988,fu2004,ohmi2007,lireho2009}. Therefore we
also tested the $p$-free disk model (i.e, $T\propto r^{-p}$, with $p$
as a free parameter, \textit{diskpbb} in XSPEC) for the three
brightest spectra X1, C1, and X2. We found that the $p$ parameter was
not well constrained for all spectra. In order to provide a meaningful
comparison of the $p$ values among different spectra, we fixed $p$ at
the value of 0.75 for a standard disk for X2 and fixed
$N_\mathrm{H,i}$ and the \textit{diskpbb} normalization (thus the
inner disk radius) at the best-fitting values obtained from the
simultaneous fit to all three spectra with these two parameters tied
together. The fit results are given in Supplementary
Table~\ref{tbl:spfit2}. We inferred $p=0.70\pm0.03$ (90\% error) for
C1, fairly close to the standard disk value (0.75) assumed in X2 (the
difference is $3\sigma$). The $p$ value from X1 ($0.610\pm0.012$) is
significantly smaller than the standard disk value (at the 20$\sigma$
confidence level). Therefore the fits with the absorbed
\textit{diskpbb} model also suggest that the inner disk reached the
local Eddington limit in X1. We note that although
  the \textit{diskpbb} model might describe X1 better
  than \textit{diskbb}, the bolometric luminosity is better to be
  estimated with \textit{diskbb} because the local Eddington limit was
  expected to be reached only in the inner disk region
  and \textit{diskpbb} could significantly overestimate emission from
  the outer disk region.

We used the AGN spectral model \textit{optxagnf}\cite{dodaji2012} (in
XSPEC) to roughly estimate the BH mass. The model was created assuming
a disk inclination of $60^\circ$. It includes three
components: a thermal disk, a low-temperature Comptonized component
and a high-temperature Comptonized component. We applied only the thermal
disk component of the model to the two thermal-state spectra that
have the highest quality, i.e., C1 and X2. Therefore, there were only
three free parameters of the model in our fits: the BH mass
$M_\mathrm{BH}$, the spin parameter $a^*$, and the Eddington ratio.
We fitted both spectra simultaneously, with all parameters except the
Eddington ratio tied to be the same. The fits could not constrain both
$M_\mathrm{BH}$ and $a^*$ well simultaneously. We found that we
required $a^*>0.92$ if C1 and X2 were required to have the Eddington
ratios below 1.0, thus in the thermal state. Assuming $a^*=0.92$, we
inferred $M_\mathrm{BH}=5.3\times10^4$ \msun\ (Supplementary
Table~\ref{tbl:spfit2}), which is likely the lower bound of the BH
mass. The upper bound of the BH mass could be obtained assuming a
maximally-rotating Kerr BH, and we got $M_\mathrm{BH}=1.2\times10^5$
\msun\ (Supplementary Table~\ref{tbl:spfit2}). We did not use
  X1 to constrain the BH mass because it is less clear what Eddington
  ratio this observation assumed, though when the inner disk
  reached the local Eddington limit, the Eddington ratio might
  be\cite{fu2004} close to 1.0.

The very soft X-ray spectra of J2150$-$0551 resemble the emission from
cooling NSs in LMXBs right after large accretion
outbursts\cite{widepa2017}. Therefore we also fitted the X-ray spectra
of J2150$-$0551 with the NS atmosphere model
\textit{nsatmos}\cite{heryna2006} (the other popular model
\textit{nsa}\cite{zapash1996} gave very similar results). We first
fixed the NS mass and radius to canonical values of
$M_\mathrm{NS}=1.4$ \msun\ and $R_\mathrm{NS}=10$ km. The X1, C1, X2,
and C2 spectra were fitted simultaneously, with the column density and
the source distance tied to be the same. The fit results are shown in
Supplementary Figure~\ref{fig:srcspec_nsatmos_withdelchi} and given in
Supplementary Table~\ref{tbl:spfit2}. The fits were marginally
acceptable. The fit to C2 had the highest $\chi^2_\nu$ value (3.2,
$\nu=8$), showing clear systematic residuals. The $\chi^2$ probability
$P_\chi$ was rather low (0.1\%). If we allowed the distance in C2 to
be different from that of X1, C1, and X2, we obtained a better fit
($\chi^2_\nu=1.9$, $\nu=7$, $P_\chi=6\%$, the fit quality was similar
to that obtained with an absorbed \textit{diskbb} model). The inferred
distance was 2.6 times smaller than that from the joint fit of X1, C1,
and X2, corresponding to a 2.5$\sigma$ discrepancy. Because it does
not make sense to adopt different source distances for different
observations, from now on, we will concentrate on the fits with the
source distance tied to be the same for all observations. The fits
inferred the effective temperature measured by a distant observer
$kT_\mathrm{nsatmos,eff}^\infty$ to decrease from 75 eV in X1 to 37 eV
in C2. We note that these effective temperatures are much lower than
the temperatures inferred from \textit{blackbody} or \textit{diskbb}
fits because the NS quiescent spectrum as calculated
in \textit{nsatmos} tends to have a harder high-energy tail than a
blackbody\cite{brbiru1998}. We could not effectively differentiate
these various models in the spectral fits because the X-ray spectra
were very soft, while we could only use energy bands above 0.3 keV and
there was Galactic absorption ($N_\mathrm{H}=3\times10^{20}$
cm$^{-2}$) at low energies.

We explored the dependence of the fits on the assumed NS mass
(1.0--2.2 \msun) and radius (8--16 km). We found that
$kT_\mathrm{nsatmos,eff}^\infty$ hardly changed ($\lesssim$4 eV) with
the assumed NS mass and radius. Supplementary Table~\ref{tbl:spfit2} also includes
the fit results for $R_\mathrm{NS}=13$ km and 16 km and
$M_\mathrm{NS}=1.4$ \msun.
\\

\noindent \textbf{The Long-term and Short-term X-ray Variability}

Figure~\ref{fig:lumlcsp} plots the long-term evolution of the
bolometric disk luminosity $L_{\rm disk}$ of J2150$-$0551 inferred from the
\textit{diskbb} fits, adopting a distance of $D_L=247$ Mpc and a disk
inclination of $60^\circ$. The source was first detected in X1 on 5
May 2006, with $L_{\rm disk}\sim1.1\times10^{43}$ erg s$^{-1}$. It was
still bright in C1, X2 and C2, but with $L_{\rm disk}$ decreasing by
factors of 1.5, 3.9, and 10.1, respectively. The source rest-frame
0.32--10.6 keV unabsorbed luminosity decreased by a factor of 17 from
X1 to C2 (Table~\ref{tbl:spfit}). The S1 observation in September 2014
had very low statistics, with only 7 counts detected (corresponding to
$\sim$3$\sigma$, one background count expected)) within 0.3--10 keV in
the source region. Therefore we did not carry out spectral fits to
S1. With 6 out of the 7 counts below 1 keV, the S1 spectrum could be
as ultrasoft as C2. Therefore we estimated the luminosity of the
source in S1 (as shown in Figure~2) assuming the \textit{diskbb} fit
to C2 and found a luminosity level consistent with that of C2.

We note that J2150$-$0551 was not detected in the {\it ROSAT} All-Sky
Survey in 1990, whose detection limit was a factor of two lower than
the flux of our source in X1. The source was covered but not detected
in an \emph{XMM-Newton} slew observation on 14 May 2004, which had an
effective exposure of 3 seconds and the $3\sigma$ detection limit of
$3\times10^{-12}$ erg s$^{-1}$ cm$^{-2}$ (a factor of four higher than
the flux in X1). The source was not detected in another
\emph{XMM-Newton} slew observation on 23 November 2015 either. This
observation had an effective exposure of 10 seconds and the detection
limit an order of magnitude above the source flux indicated by S1 and
C2.

Supplementary Figure~\ref{fig:asmlightcurve} shows the long-term
monthly light curve of J2150$-$0551 from the ASM onboard
the \emph{RXTE} over the whole time of the mission. There was no
significant detection of the source in any timescale. On the monthly
basis, the $3\sigma$ upper limit when the source
flux was not significantly affected by the Sun was typically $\lesssim$0.2 cts
s$^{-1}$, corresponding to $<$$6\times10^{-11}$ erg s$^{-1}$ cm$^{-2}$
(1.5--12 keV). This limit is three orders of magnitude more than the peak
 flux of J2150$-$0551 in 1.5--12 keV ($5\times10^{-14}$ erg s$^{-1}$
 cm$^{-2}$, from X1), explaining why the ASM did not detect the soft
 X-ray outburst detected by \emph{XMM-Newton} and \emph{Chandra}.

Supplementary Figure~\ref{fig:srclc} shows the light curves of J2150$-$0551 from the
four deep X-ray observations X1, C1, X2, and C2. We obtained $\chi^2$
probabilities of constancy of 0.07, 1.0, 0.02, and 0.48 for these light
curves, respectively, indicating no significant short-term
variability in all observations.\\

\noindent \textbf{Multi-wavelength Photometry Observations}

The photometry of J2150$-$0551 in various filters is listed in
Supplementary Table~\ref{tbl:obslogopt} and plotted in Supplementary
Figure~\ref{fig:srcspec_diskirspp}.  The source appeared red in the
CFHT/CFH12K images in 2000--2001 but became blue in the CFHT/MegaCam
images in 2005. It seemed to become red again in our new SOAR, Subaru,
and Gemini images in 2014--2016. The change was thus more significant
at shorter wavelengths, by $\sim$0.8 mag in the $g\arcmin$ band and
$\sim$0.3 mag in the $r\arcmin$ band. The {\it HST}/ACS F775W
photometry in September 2003 was consistent with the CFHT/CFH12K
photometry, but not with the CFHT/MegaCam photometry. Therefore, the
optical variability indicates that the outburst activity of
J2150$-$0551 began before June 2005 (i.e., about one year before X1 or
earlier) but after September 2003.

The CFHT/CFH12K photometry would then represent the quiescent
emission of the optical counterpart to J2150$-$0551. The absolute
$V$-band magnitude of the source would be
$\sim$$-12.3$ AB mag if it is in Gal1. Therefore
J2150$-$0551 most likely resides in a massive star cluster. In order
to assess the properties of the cluster, we fitted the CFHT/CFH12K and
{\it HST}/ACS photometry with a stellar population model\cite{ma2005}
that is based on theoretical atmospheres with the Salpeter initial
mass function. We applied both Galactic dust reddening of
$\mathrm{E(B-V)}_\mathrm{G}=0.03$ mag and the intrinsic reddening of
$\mathrm{E(B-V)}_\mathrm{i}=0.01$ mag. The intrinsic reddening was
calculated based on the relation $\mathrm{E(B-V)}=1.7\times
10^{-22}N_\mathrm{H}$ and using intrinsic
$N_\mathrm{H}$ obtained from the X-ray spectral fits. The redshift of
$z=0.055$ was applied. We inferred a stellar population of age
$1.0\pm0.3$ Gyr and bolometric luminosity
$L_*=1.4\times10^7$ \lsun, assuming a solar metallicity. Using the
mass-to-light ratio for the corresponding metallicity and
age\footnote{http:\/\/www.maraston.eu}, we estimated the stellar mass
of the cluster to be
$M_*=8.0\times10^6$ \msun. We also tested a two-population solution
with the age of one population fixed at 14 Gyr. We found that we could
not constrain the individual populations well and could not rule out a
cluster composed of a dominant old population
($M_*\sim5\times10^7$ \msun) and a small young population
($M_*\sim4\times10^5$ \msun). Finally, we checked the fit with a metal
poor population of metallicity
[Z/H]=$-1.35$ and inferred a population of age
$10_{-4}^{+5}$ Gyr,
$M_*\sim2.9\times10^7$ \msun, and
$L_*\sim1.3\times10^7$ \lsun. The various fits suggest the cluster
to have
$M_*\sim(0.8$--$5.0)\times10^7$ \msun\ and
$L_*\sim1.4\times10^7$ \lsun. The inferred population age has
  large uncertainties, strongly dependent on the population model
  assumed.

The enhancement of the flux at short optical wavelengths in 2005 could
be from an irradiated accretion disk near the peak of the outburst,
while radiation-driven outflow/wind could also cause this. In order to
check roughly whether the optical flare could be described with disk
irradiation, we fitted the broad-band spectra combining the X1 X-ray
spectrum and the CFHT/MegaCam and CFHT/WIRCam photometry with the
irradiated accretion disk model \textit{diskir}\cite{gidopa2009} (in
XSPEC), with an additional component to account for the star cluster
emission. The star cluster component was fixed at the single stellar
population model based on the CFHT/CFH12K and {\it HST}/ACS
photometry, obtained above with a solar metallicity. The irradiated
accretion disk model included thermal emission from the inner disk and
reprocessing in the outer disk, but not the Compton tail, as the X-ray
spectrum seemed purely thermal. The best-fitting model is shown in
Supplementary Figure~\ref{fig:srcspec_diskirspp}. The fraction of the
bolometric flux that was thermalized in the outer disk was inferred to
be $f_\mathrm{out}=1.4_{-0.4}^{+1.6}\times10^{-3}$, which is close to
the typical values of $\sim$$10^{-3}$ seen in BH X-ray
binaries\cite{gidopa2009}. The best-fitting outer disk radius
$r_\mathrm{out}$ was $10^{3.9\pm0.3}$ times the inner disk radius. We
note that the CFHT/MegaCam photometry was taken before X1 and could
correspond to a higher X-ray flux and that there could be hot outflow
which we neglected in the modeling. Then we expect the real disk to
assume smaller values of $r_\mathrm{out}$ and/or $f_\mathrm{out}$ than
obtained above.

The model obtained from the broad-band fit above predicted the flux in
the OM $UVM2$ filter to be one order of magnitude lower than the
$3\sigma$ detection limit of the OM $UVM2$ images in X1 and X2
($\sim$22.0 AB mag), explaining the non-detection of our source in
these images. The $3\sigma$ detection limit (24.2 AB mag) of the
$UVW2$ image in S1 is a factor of $\sim$2 above the prediction from
the broad-band fit. Further considering the significant decrease of
the source X-ray flux from X1 to S1, we expect the $UVW2$ flux in S1
to be much lower than the detection limit, explaining the
non-detection of our source in the S1 $UVW2$ image. \\

\noindent \textbf{The GMOS Observation}

Supplementary Figure~\ref{fig:geminispfit_Gal1} plots the Gemini
spectrum of the nuclear region of Gal1. The spectrum showed prominent
absorption features, and thus we expect it to be dominated by an old
stellar population. This is supported by the pPXF fit, which inferred
the mean age weighted by the mass to be 10.9 Gyr and the mean age
weighted by the light to be 10.4 Gyr (Supplementary
Figure~\ref{fig:masslightdis_Gal1}). Some faint emission lines were
detected too, most significantly [\ion{N}{II}] $\lambda$6583
(luminosity $L=9\pm1\times10^{39}$ erg s$^{-1}$, 1$\sigma$ error;
[\ion{N}{II}] $\lambda$6547 was affected by the atmospheric OH
absorption) and the [\ion{S}{II}] doublet (total
$L=9\pm1\times10^{39}$ erg s$^{-1}$). H$\alpha$ were weaker, with
$L=5\pm1\times10^{39}$ erg s$^{-1}$. The emission lines in the
H$\beta$-[\ion{O}{III}] region, if present, were very weak (both
[\ion{O}{III}] $\lambda$4969 and $\lambda$5007 had
$L=3\pm1\times10^{39}$ erg s$^{-1}$ and H$\beta$ had
$L<3\times10^{39}$ erg s$^{-1}$, $3\sigma$ upper limit). The emission
lines are probably due to nuclear activity, because [\ion{N}{II}]
$\lambda$6583 was stronger than H$\alpha$ and we also detected a
nuclear hard X-ray source (Source 1). The pPXF fit inferred the total
stellar mass of Gal1 to be $2.3\times10^{11}$~\msun\ and the total
stellar light within the fitting band (rest-frame 4400--7000 \AA) to
be $1.5\times10^{10}$ \lsun, which has been corrected for the slit
loss by comparing the spectrum with the SDSS $r\arcmin$-band
photometry.

The spectrum of Gal2 is shown in Supplementary
Figure~\ref{fig:geminispfit_G2}, and it is somewhat noisy. There were
no significant emission lines or absorption features that would allow
us to determine the redshift of the galaxies. In the pPXF fit shown in
figure, we assumed Gal2 to have the same redshift as Gal1. The
spectrum appeared blue, and our pPXF fit suggests the presence of
young stellar populations ($\lesssim0.3$ Gyr, Supplementary
Figure~\ref{fig:masslightdis_G2}). The mean age of Gal2 weighted by
the mass is 7.6 Gyr, while the mean age weighted by the light is 2.5
Gyr. The total mass is $1.1\times10^8$ \msun, and the luminosity
within the fitting band is $4.7\times10^7$ \lsun.

The spectrum of the optical counterpart to J2150$-$0551 is even
noisier. There were no clear emission lines or absorption features in
this spectrum either. In the pPXF fit shown in Supplementary
Figure~\ref{fig:geminispfit_J2150$-$0551}, we assumed J2150$-$0551 to
be blueshifted by 300 km s$^{-1}$ relative to Gal1 so that the
H$\alpha$ line corresponded to a spike in the spectrum. In this case,
there might be some H$\alpha$ emission at $L=2.1\pm0.4\times10^{38}$
erg s$^{-1}$. The absence of strong narrow emission lines in
  J2150$-$0551 could be due to the lack of a narrow-line region as
  seen in AGNs and/or due to very weak persistent emission. The
  $3\sigma$ upper limit for [\ion{O}{III}] $\lambda$5007 was
  $2\times10^{37}$ erg s$^{-1}$.  Based on the
  relation\cite{labima2009} between the bolometric persistent
  luminosity and the [\ion{O}{III}] $\lambda$5007 luminosity for AGNs,
  we inferred the bolometric persistent luminosity in J2150$-$0551 to
  be $<$$1.7\times10^{39}$ erg s$^{-1}$
  ($3\sigma$ upper limit), which is a factor of
  $\sim$6,000 below the peak luminosity of J2150$-$0551.

Because the spectrum is too noisy, the mass and light distributions inferred from the pPXF fit probably have large uncertainties. Thus we do not show them here. The pPXF fit inferred the optical counterpart to J2150$-$0551 to have
a total stellar mass of $4.9\times10^7$~\msun, and stellar light
within the fitting band of $5.7\times10^6$~\lsun. The mean age
weighted by the mass is 7.8 Gyr, and the mean age weighted by the
light is 5.8 Gyr. These values are broadly consistent with those
obtained from the fit to the CFHT/CFH12K and \emph{HST} 2000--2003
photometry with a single stellar population model. In Supplementary
Figure~\ref{fig:geminispfit_J2150$-$0551}, we also mark the positions
of the typical emission/absorption features expected if the source is
a Galactic object of zero redshift. We could not confidently identify
any feature in this case either. \\

\noindent \textbf{The off-center IMBH Explanation}

We have fitted the X-ray spectra of J2150$-$0551 assuming it to be in
Gal1. All the results are well explained if it is a TDE from an IMBH
of mass $\sim$$5\times10^4$ \msun. The IMBH nature is strongly
supported because the X-ray spectra can be described with a standard
thermal thin disk of very high luminosities and low temperatures,
which approximately follow the $L\propto T^4$ relation. Such a
relation, though commonly observed in accreting stellar-mass BH in the
thermal state\cite{remc2006,dogiku2007}, was not clearly observed in
other IMBH candidates except\cite{sefali2011,goplka2012} ESO 243-49
HLX-1. The position of the $L\propto T^4$ relation in the $L$--$T$
plot and thus the disk temperature range (0.14--0.28 keV in
J2150$-$0551) depend mainly on the BH mass. Therefore the
similar $L\propto T^4$ relations traced out by ESO 243-49 HLX-1 and
J2150$-$0551 (Figure~3) suggest that they contain BHs of
similar masses, within a factor of a few.

The IMBH nature of J2150$-$0551 can also be inferred based on
comparison of its disk temperatures with those seen in TDEs from
galactic nuclei. Nuclear TDEs with good thermal X-ray spectra have
peak disk temperatures in the
range\cite{ko2002,licagr2011,limair2015,liguko2017,mi2015} between
0.05--0.15 keV (for studies that used a single-temperature blackbody
to fit the X-ray spectra, we have multiplied the blackbody temperature
by a factor of 1.4 to infer the equivalent disk
temperature\cite{mamami1986}). The peak disk temperature of
J2150$-$0551 (0.28 keV) was at least twice higher than this
range. Because the disk temperature depends on the BH mass as $M^{-1/4}$ for a standard thermal thin
disk\cite{shsy1973}, the mass of the BH in J2150$-$0551 would
then be at least one order of magnitude lower than seen in most nuclear TDEs
and thus is most likely in the IMBH range.

We note that  J2150$-$0551 is unlikely to be one of ultraluminous X-ray
pulsars, whose luminosities have been detected\cite{isbest2017} to be up
to $10^{41}$ erg s$^{-1}$. The ultrasoft X-ray spectra (photon index
$\ge$4.8) of J2150$-$0551
are in contrast with the very hard spectra (photo index $<$2.0 below 5
keV) of four known ultraluminous X-ray
pulsars\cite{isbest2017,wafuha2018}.

Our multiwavelength data suggest that J2150$-$0551 exhibited
  a luminous prolonged outburst from optical to soft X-rays. The outburst
  was probably not caused by large variability of persistent emission,
  which might be very weak based on the lack of strong narrow
  emission lines (see the previous section). The high luminosity and
  the long duration of the outburst can be naturally explained if the
  event is a TDE. A simple TDE model for the luminosity evolution of
  J2150$-$0551 is plotted in Figure 2. It models the decay as
  $(t-t_\mathrm{D})^{-5/3}$, which was predicted by the standard TDE
  theory\cite{re1988,ph1989} and observed in many TDE candidates in
  galactic nuclei\cite{ko2015,hugebl2017}. The disruption time
  $t_\mathrm{D}$ was inferred to be around 18 October 2003 (1$\sigma$
  error 37 days). The very small number of data points that we had did
  not allow us to fit the light curve with more complicated models. If
  we fitted the luminosity decay as $(t-t_\mathrm{D})^{\alpha}$ with a
  variable index $\alpha$, we obtained a fit of $\alpha=-0.8\pm0.1$
  (1$\sigma$ error), but via the $F$-test, a variable index only
  improved the fit at the 74\% confidence level. Moreover, the
  disruption time was inferred to be on 24 August 2005 (1$\sigma$
  error 47 days), within the time interval when the optical flare was
  detected. Similarly, with so few data points, it is difficult to
  check for the presence of slow circularization.  When
  circularization of stellar debris is slow, the light curve would
  decline slower than predicted in the standard TDE
  theory\cite{gura2015}, as might occur in the TDE
  candidate\cite{liguko2017} 3XMM~J150052.0+015452. Given that the
  luminosity evolution of J2150$-$0551 is consistent with the standard
  $(t-t_\mathrm{D})^{-5/3}$ curve, we expect relatively prompt
  circularization of stellar debris in this event. 

There could be a super-Eddington accretion phase in J2150$-$0551
lasting for around three years. This is supported by the X1
observation, which seems to deviate from the $L\propto T^4$ relation
traced out by the other observations. The luminosity in the
super-Eddington accretion phase is expected to be around the Eddington
limit\cite{krpi2012}. This can explain the non-detection of the source
in the \emph{XMM-Newton} slew observation in 2004 (Figure~2). In the
optical flare phase detected in 2005, the $u^*$-band flux in June 2005
was close to that expected from the $g\arcmin$ and $r\arcmin$ fluxes
in November 2005, which also supports a constant luminosity in the
super-Eddington accretion phase, while the mass accretion rate was
expected to decrease by 50\% in this period (Figure~2). The
super-Eddington accretion phase can last\cite{stsalo2013} $t_{\rm
Edd}\approx 1.9(\eta/0.1)^{3/5}M_6^{-2/5}m_*^{1/5}x_*^{3/5}$ yr, where
$M_6=M_{\rm BH}/10^6$ \msun, $\eta$ is the radiative efficiency, $m_*$
is the mass of the star in units of \msun, and $x_*$ is the radius of
the star in units of \rsun. Assuming typical parameters of $\eta=0.1$,
$m_*=1.0$, and $x_*=1.0$, we have $t_\mathrm{Edd}=12.0$ yr and 4.8 yr
for $M_{\rm BH}=10^4$ \msun\ and $10^5$ \msun, respectively. Therefore
our inference of $t_{\rm Edd}\sim 3$ yr for J2150$-$0551 is reasonable
for its IMBH nature. Because of the presence of a long super-Eddington
accretion phase, which was largely missed, J2150$-$0551 appeared to
last longer than other TDEs (typically decay significantly in a year).

The optical counterpart in quiescence is consistent with a massive
($\sim$10$^7$ \msun) star cluster with an effective radius of $<$20
pc. This dense star cluster environment makes the TDE explanation for
J2150$-$0551 reasonable. One intriguing explanation for the optical
counterpart is that it is the accreted remnant nucleus of a dwarf
galaxy tidally stripped by Gal1. Accreted remnant nuclei of dwarf
galaxies have been used to explain massive star clusters brighter than
classical globular
clusters\cite{hiinvi1999,drjogr2000,phdrgr2001,begoqu2010}. Interestingly,
Figure~\ref{fig:hstimg} shows a possible minor merger of a satellite
galaxy (Gal2), which is connected with Gal1 through a possible stream
and is close to J2150$-$0551. The nuclei are normally expected to
remain in the middle of the tidal stream in minor
mergers\cite{jerobr2015}. Thus Gal2 is probably unrelated to
J2150$-$0551. However, the presence of such a merging feature
indicates that Gal1 might be in an epoch of frequent minor mergers.

A few optical transients possibly offset from the centers of their
host galaxies were reported as TDE candidates (PTF-10iam, PTF-10nuj,
PTF-11glr, and Dougie)\cite{argasu2014,viyuqu2015}, but it is very
hard to completely rule out a supernova explanation for
them\cite{arwoho2016}, due to the lack of strong X-ray detections
associated with these events\cite{augura2017}. The event 3XMM
J141711.1+522541 could be an off-center TDE with a peak X-ray
luminosity of a few $\sim$$10^{43}$ erg
s$^{-1}$, but the TDE nature of that event was not certain due to poor
sampling of the light curve\cite{licawe2016}. Therefore, J2150$-$0551
is the best off-center TDE candidate thus far.

A wide variety of searching
strategies\cite{reco2016,kafero2017,me2017} have resulted in discovery
of a few
off-center\cite{geriho2005,faweba2009,sefali2011,goplka2012,wecsle2012,lukige2013,pastmu2014,lugeba2015,merolo2015,licawe2016,kibalo2017}
and
centered\cite{bahoru2004,hokite2012,maliir2014,maulro2014,barega2015}
IMBH candidates.  These objects were generally not bright in X-rays,
except ESO 243-49 HLX-1 and some others that could be due to
TDEs\cite{mauler2013,maliir2014,licawe2016,ligoho2017}. ESO 243-49
HLX-1 and J2150$-$0551 showed the best evidence for the thermal state of
an IMBH, with a clear $L\propto T^4$ evolution track detected.
\\

\noindent \textbf{Alternative Explanations}

At a high Galactic latitude of $-42^{\circ}$, J2150$-$0551 has a high
chance to be extragalactic. We have explored the case that the source
is in Gal1. We still need to check whether the source could be instead
from the nucleus of a background galaxy, either as an AGN or TDE, that
just happens to lie very close to Gal1 in the sky.  J2150$-$0551 is
different from standard AGNs in various aspects, with very soft X-ray
spectra ($\Gamma_{\rm PL}>4.5$), large variability (a factor of
$\sim$17, absorbed 0.2--12 flux), and a large X-ray (0.2--12 keV,
maximum) to IR ($K_\mathrm{s}$-band) flux ratio
($\log(F_{\rm X}/F_{\rm IR})=3.2$), while most AGNs, including
  Narrow-line Seyfert 1 galaxies that might have accretion rates close
  to the Eddington limit, have\cite{liweba2012}
$\Gamma_{\rm PL}\sim1.9$, long-term variation factors $<10$, and
$\log(F_{\rm X}/F_{\rm IR})\lesssim2.0$.  Two AGN candidates (2XMM
J123103.2+110648 and GSN 069) with AGN emission line signatures in
optical were found to have pure thermal X-ray spectra and large
variability, but the TDE explanation for them cannot be ruled
out\cite{liweba2012,liirgo2013,liweba2014,tekaaw2012,hokite2012,misaro2013}. One
main difference between J2150$-$0551 and these two and other thermal
TDE candidates is that J2150$-$0551 had a significantly higher
X-ray-to-IR flux ratio ($\log(F_{\rm X}/F_{\rm IR})=3.2$) than other
thermal TDE candidates ($\log(F_{\rm X}/F_{\rm
  IR})\lesssim2.0$). Moreover, there are very few known thermal TDEs,
especially those of high characteristic temperatures (peak temperature
$>0.2$ keV). Then the chance probability to find one such TDE so close
to Gal1 is very low
($\sim$$0.01$\%). Based on the above argments, we rule out a
background AGN or TDE explanation for J2150$-$0551.

If J2150$-$0551 is a Galactic object, given its high Galactic
latitude, we expect the source distance to be small. Then it would
most likely have low luminosities (0.2--12 keV luminosity
$<$$5\times10^{32}$ erg s$^{-1}$ assuming a distance of 2
kpc). J2150$-$0551 had $\log(F_{\rm X}/F_{\rm IR})=3.2$, making it
unlikely to be a coronally active star ($\log(F_{\rm X}/F_{\rm
IR})\lesssim-0.9$)\cite{liweba2012}. High X-ray to IR flux
ratios, however, are often seen in compact object systems that contain
white dwarfs, NSs, or BHs\cite{liweba2012}. With very soft X-ray
spectra, the source cannot be a low-luminosity BH X-ray binary, whose
X-ray spectra are expected to be hard
($\Gamma_\mathrm{PL}\lesssim2.0$)\cite{plmiga2017}. The large
variability (a factor of 17 in $\sim$10 yr) of the source makes it
unlikely to be an isolated white dwarf or NS. The low luminosities of
the source would make it unlikely to be supersoft emission from
cataclysmic variables ($\gtrsim10^{36}$ erg s$^{-1}$)\cite{kava2006}.

Then we are left with the cooling NS in a LMXB as the only likely
alternative explanation for the source if it is in our Galaxy. As we
have shown above, the very soft X-ray spectra of J2150$-$0551
approximately resemble those of thermally cooling NSs in LMXBs often
detected after large accretion outbursts\cite{deoore2017}. Supplementary
Figure~\ref{fig:coolingcurve} shows that the cooling curve of
J2150$-$0551 (if it is an NS LMXB) is very similar to those of six
confirmed LMXBs\cite{hofrwi2014}. There are two major differences
though. One is that J2150$-$0551 had a cooling timescale of 2008
days, much longer than the six NS LMXBs shown ($<$500
days). The other major difference is that the NS in J2150$-$0551 in
the later stage would be much cooler (31 eV) than the NSs in those six
LMXBs ($>$50 eV).

In the NS LMXB explanation for J2150$-$0551, the very faint optical
flare detected in the CFHT images in 2005 would be probably associated
with the accretion outburst, instead of the early fast-evolving
thermal cooling phase. This is because the optical flux was steady
during the flaring period (see the above section). Then we inferred
the accretion outburst to last for at least five months and end at a
time between 5 May 2006 (when X1 was taken) and 7 November 2005 (when
the CFHT $g'$-band image was taken). The accretion outburst should
have started after 14 May 2004 (when the first \emph{XMM-Newton} slew
observation was taken). This is because J2150$-$0551 was not detected
in this slew observation, whose $3\sigma$ upper limit of the 0.2--12
keV flux was only a factor of four of the flux in X1. Our constraint
on the accretion outburst start date is also supported by the absence
of an optical enhancement in the \emph{HST} observation in September
2003.

The quiescent optical/IR counterpart to J2150$-$0551 is very faint,
with the $K_\mathrm{s}$-band Vega magnitude of at most 20.8 mag, or
absolute magnitude between +8.1 and +9.1 mag assuming a source
distance of 2.2--3.5 kpc (inferred from the \textit{nsatmos} fits
listed in Table~\ref{tbl:spfit2}). The companion counterpart is
then expected to be a late M dwarf\cite{duli2012}. However, during quiescence, the
optical spectrum of the source did not look very red, as would be
expected for a late M dwarf. The most likely explanation for this
would be that the companion was illuminated by the NS thermal
emission.

Besides the very low chance probability of positional coincidence with
Gal1, the most challenging problem with the cooling NS LMXB explanation
for J2150$-$0551 is the absence of a large accretion outburst in the
ASM light curve, which constrained the $3\sigma$ upper limit of the
monthly-average 1.5--12 keV luminosity of J2150$-$0551 to be
(3--9)$\times10^{34}$ erg s$^{-1}$, assuming a source distance of
2.2--3.5 kpc (Table~\ref{tbl:spfit2}). Then the accretion outburst
would be very faint, making the source one of the very faint X-ray
transients (VFXTs, peak X-ray luminosity $<$$10^{36}$ erg s$^{-1}$),
most of which were detected toward the Galactic
center\cite{wiinru2006,dewi2009,ca2009}. Most VFXTs have very short
outbursts of weeks, but quasi-persistent VFXTs also
exist\cite{hebade2015}. Cooling curves thus far were only detected for
transient LMXBs that experience bright outbursts ($\gtrsim$$10^{36}$
erg s$^{-1}$). It is not clear whether the crust of an NS in a VFXT could be
significantly heated up.

Our numerical modeling with \texttt{NSCool} of the hypothetical
accretion outburst that may have resulted in the observed cooling
curve of an NS in the LMXB J2150$-$0551 inferred the mass accretion
rate $\dot M$ and thus ASM count rate $\dot C$ to strongly depend on
the value of $Q_\mathrm{sh}$ assumed. The physical nature of the
shallow heating is still unknown, and $Q_\mathrm{sh}$ inferred by
fitting cooling curves of several LMXBs are
mostly\cite{brcu2009,pare2013,oopawi2016,mecabr2016,wadewi2016} $<$3
MeV/nucleon. The MCMC runs found $\dot C=2.3_{-0.5}^{+0.8}$ cts
s$^{-1}$ ($1\sigma$ error) for $Q_\mathrm{sh}=1$ MeV and $\dot
C=0.8_{-0.2}^{+0.3}$ cts s$^{-1}$ for $Q_\mathrm{sh}=3$ MeV. These
values are factors of 10 ($11\sigma$) and 4 ($5\sigma$) of the
$3\sigma$ upper limit implied by the ASM ($\lesssim$0.2 cts
s$^{-1}$), respectively. The discrepancy would be larger, further
considering that our MCMC runs assumed a constant $\dot M$ in the
accretion outburst, without taking into account the outburst decay,
which would significantly underestimate\cite{oopawi2016}
$Q_\mathrm{sh}$ and thus $\dot M$. If $Q_\mathrm{sh}$ was allowed to
be extremely high (10 MeV/nucleon), it is possible to have a very
faint accretion outburst ($\dot C=0.21\pm0.04$ cts s$^{-1}$, around
the ASM detection limit) to produce the observed cooling
curve. However, thus far, only MAXI J0556$-$332, which is completely
different from J2150$-$0551 for containing an extremely hot NS (the
top dotted line in Supplementary Figure~\ref{fig:coolingcurve}) and
having super-Eddington accretion rates in the peak\cite{hofrwi2014},
was inferred to have\cite{decubr2015,pahowi2017} $Q_\mathrm{sh}\sim10$
MeV/nucleon.

One way to differentiate between a Galactic cooling NS and a TDE by an
extragalactic IMBH would be through continued multiwavelength
follow-up observations. NS LMXBs in quiescence might sporadically exhibit
low-luminosity X-ray flares\cite{frhowi2011,hofrwi2014},
which are not expected in a TDE. If it is a cooling NS, we expect the
flux and the temperature to hardly decrease in the future, while for a
TDE, we expect the flux to continue to decrease with time,
approximately as $t^{-5/3}$.\\

\noindent \textbf{IMBH Space Density}

TDE rates can be used to probe the space density and the occupation
fraction of BHs\cite{filo2017}. We roughly estimated the rate
of off-center TDEs like J2150$-$0551, in a similar way as we did for
another TDE candidate\cite{liguko2017} 3XMM~J150052.0+015452. We
discovered J2150$-$0551 through systematic search over the 3XMM-DR5
catalog\cite{rowewa2016}. We were limited to the sources that have at least one
detection of S/N $>$ 20. We focused on sources of large variability
and/or soft spectra and carefully inspected whether they are TDEs
based on all available public data. Limited to observations outside
the Galactic plane (galactic latitude $|b|>20^\circ$) and assuming
that events like J2150$-$0551 have a mean luminosity and soft X-ray
spectra as in X1 and have a duration of 3 years, we estimated to detect
$\sim$$2.0\times10^8 r$ events, where $r$ is the event rate in units
of Mpc$^{-3}$ yr$^{-1}$. We at least discovered one event, suggesting
$r\gtrsim 5\times10^{-9}$ Mpc$^{-3}$ yr$^{-1}$. Our search was not
complete in the sense that we could not confirm the off-center TDE
nature of a few fainter events due to the poor X-ray coverage and/or
the lack of deep optical data (needed to check the off-center nature
of the events). For example, another event 3XMM J141711.1+522541
possibly similar to J2150$-$0551 was also discovered in this
search\cite{licawe2016}. Taking this into account, we estimated $r$ to
be around $10^{-8}$ Mpc$^{-3}$ yr$^{-1}$.

The expected TDE rate $\gamma$ of IMBHs of a few $10^4$ \msun\ in star
 clusters is not well established theoretically.  Studies with direct
 $N$-body simulations\cite{bamaeb2004,brbakr2011} inferred
 $\gamma\sim10^{-5}$ yr$^{-1}$ per IMBH of a few $10^4$ \msun, while
 analytical studies
 estimated\cite{wame2004,ko2016,stme2016}
 $\gamma\sim10^{-3}$ yr$^{-1}$. Then based on the observed TDE rate
 above, we constrained the space density $n$ of off-center IMBHs of a
 few $10^4$ \msun\ to be between $\sim$$10^{-5}$--$10^{-3}$
 Mpc$^{-3}$.

 The space density
 $n\arcmin$ of the off-center star clusters of mass
 $\ge$$10^7$ \msun, as observed in the counterpart to J2150$-$0551,
 can be estimated based on the dependence of the number of massive
 star clusters on the halo mass predicted from numerical
 simulations\cite{pfgrba2014} and the halo mass
 function\cite{muporo2013}. We estimated $n\arcmin\sim10^{-2}$
 Mpc$^{-3}$, which had been multiplied by a factor of 2 to account for
 the fact that the numerical simulations appeared to underestimate the
 number of massive star clusters in galaxy clusters by such a
 factor. Therefore, the space density of massive star clusters is, as
 expected, at least one order of magnitude larger than the space
 density of IMBHs of a few $10^4$ \msun\ estimated by us above based
 on the TDE rate.

\clearpage

\begin{table*}
\centering
\caption{\textbf{The X-ray Observation Log}. Columns: (1) the
  observation ID with our designation given in parentheses, (2) the
  observation start date, (3) the instrument, (4) the off-axis angle,
  (5) the exposures of data used in final analysis, (6) the radius of the source extraction region, (7) the net count rate in the source extraction region (0.3--10 keV for \emph{XMM-Newton} and {\it Swift} observations, and 0.4--8 keV for C1 and 0.3--8 keV for C2, all in the observer frame), with 90\% confidence error. }
\label{tbl:obslog}
\bigskip
\scriptsize
\sffamily
\begin{tabular}{rcccccccc}
\hline
\hline
Obs. ID &Date & Instrument &OAA (arcmin) &$T$ (ks) &$r_{\rm src}$ (arcsec)  & Count rate (10$^{-3}$ counts~s$^{-1}$)\\
(1) & (2) &(3) & (4) & (5) & (6) & (7)\\
\hline
\multicolumn{4}{l}{\emph{XMM-Newton}:}\\
\hline
0404190101(X1) &2006-05-05 & pn/MOS1/MOS2 & 7.1 & 24/50/50 & 35/25/25 & $115.4\pm3.8$/$68.6\pm1.9$/$63.2\pm1.9$ \\
0603590101(X2) &2009-06-07 & pn/MOS1/MOS2 & 5.8 & 52/69/69 & 35/35/35 & $97.8\pm2.5$/$22.6\pm1.0$/$22.5\pm1.0$ \\
\hline
\multicolumn{4}{l}{{\it Chandra}:}\\
\hline
6791(C1) & 2006-09-28 & ACIS-I0 &  8.1 & 101 & 7.7 & $49.0\pm1.1$\\
17862(C2) & 2016-09-14 & ACIS-S3 & 0.15 & 77 & 1.6 & $2.0\pm0.3$\\
\hline
\multicolumn{4}{l}{{\it Swift}:}\\
\hline
00033396001--2(S1) & 2014-09-05 & XRT & 2.4 & 8.0 & 20 & $0.74_{-0.45}^{+0.67}$ \\
\hline
\hline
\end{tabular}
\end{table*}

\clearpage

\begin{table*}
\centering
\caption{\textbf{The Optical/IR Observation Log and Photometry}. Columns: (1) the telescope, (2) the observation start
  date, (3) the instrument, (4) the filter, (5) effective wavelength, (6) full width at the half maximum; (7) the exposure time; (8)
 the AB magnitude with $1\sigma$ error. \textsuperscript{a}The
 exposure time is half of that of the observation, because our source
 was in the CCD gap half of the time. \textsuperscript{b}Including two
 exposures (each of 840 s) in August--September
 2001. \textsuperscript{c}Including two exposures (each of 510 s) in
 August--September 2001.}
\label{tbl:obslogopt}
\bigskip
\scriptsize
\sffamily
\begin{tabular}{rcccccccc}
\hline
\hline
Telescope & Date & Instrument &Filter &$\lambda_\mathrm{eff}$ (\AA) &$\Delta\lambda_\mathrm{FWHM}$ (\AA) &$T$ (s) &Magnitude (AB mag)\\
(1) & (2) &(3) & (4) & (5) &  (6) &(7)&(8)\\
\hline
{\it HST} &2003-09-13 & ACS & F775W &7653&1517&1020$^\mathrm{a}$  & $23.87\pm0.03$\\
CFHT & 2000-08-03& CFH12K MOSAIC& $B$ & 4302 &990& 7080$^\mathrm{b}$  & $25.27\pm0.08$\\
     & 2000-08-01 & CFH12K MOSAIC& $V$ & 5338 &974& 7440  & $24.60\pm0.06$ \\
     & 2000-08-03& CFH12K MOSAIC& $R$ & 6516 &1250& 7080$^\mathrm{c}$ & $24.07\pm0.04$\\
     &2000-08-01& CFH12K MOSAIC& $I$ & 8090 & 2164 & 21600  & $23.78\pm0.03$\\
& 2005-06-03& MegaCam & $u^*$ &3882&654& 7040  & $24.41\pm0.05$ \\
    &2005-11-07 & MegaCam & $g\arcmin$ &4767&1434& 744 & $24.03\pm0.10$\\
   &2005-11-07 & MegaCam & $r\arcmin$ &6192&1219& 480  & $23.70\pm0.07$\\
   &2005-11-07 & MegaCam & $z\arcmin$ &8824&934& 360  & $23.43\pm0.16$\\
    &2009-06-14 & WIRCam & $K_{\rm s}$ &21338&3270&900  & $22.64\pm0.10$ \\
SOAR &2014-11-26&Goodman Spectrograph&$U$& 3664 & 653 &2400&$25.97\pm0.39$\\
Subaru&2014-12-19&Suprime-Cam & $g\arcmin$ &4619 &1390 &535 & $24.78\pm0.07$\\
Gemini  &2016-08-06 &Gemini-S &$r_\mathrm{G0326}$ & 6244 &1415 & 720 & $24.04\pm0.04$\\
\hline
\hline
\end{tabular}
\end{table*}

\begin{table*}
\centering
\caption{\textbf{Fitting results of the X-ray spectra from J2150$-$0551.} All errors are at the 90\% confidence level. $N_\mathrm{diskpbb}$ is defined as $((R_\mathrm{in}/\mathrm{km})/(D/\mathrm{10 kpc}))^2\cos\theta$,
  where $R_\mathrm{in}$ is the apparent inner disk radius, $D$ is
  the source distance, $\theta$ is the disk inclination. \textsuperscript{a}These parameters were tied together in the simultaneous fits to the X1, C1, and X2 spectra, and they were fixed at the best-fitting values when we calculated uncertainties of other spectral parameters. }
\label{tbl:spfit2}
\bigskip
\scriptsize
\sffamily
\begin{tabular}{rcccccccc}
\hline
\hline
&X1 & C1 & X2 & C2\\
\hline
\multicolumn{4}{l}{Model: diskpbb}\\
$N_\mathrm{H,i}$ (10$^{20}$ cm$^{-2}$)\textsuperscript{a} & \multicolumn{3}{c}{$1.2\pm0.8$}&... \\
$kT_\mathrm{diskpbb}$ (keV) & $0.283^{+0.003}_{-0.003}$  & $0.281^{+0.004}_{-0.004}$  & $0.230^{+0.002}_{-0.002}$ &... \\
$p_\mathrm{diskpbb}$ & $0.61\pm0.01^{+0.012}_{-0.012}$  & $0.71\pm0.03$  & $0.75$(fixed)&... \\
$N_\mathrm{diskpbb}$\textsuperscript{a} & \multicolumn{3}{c}{$7.5\pm1.2$}&... \\
$\chi^2_\nu(\nu)$&$0.99(326)$&$1.28(101)$&$0.96(361)$&... \\
\hline
\multicolumn{4}{l}{Model: optxagnf ($a^*=0.92$, pure thermal disk)} \\
$N_\mathrm{H,i}$ (10$^{20}$ cm$^{-2}$)\textsuperscript{a} & ... &\multicolumn{2}{c}{$0.4^{+0.5}$}&... \\
$M_\mathrm{BH}$ (\msun)& ... & \multicolumn{2}{c}{ $ 5.3\pm0.4\times10^4$}&... \\
$L_\mathrm{bol}/L_\mathrm{Edd}$ & ... & $ 1.003\pm0.034$ & $ 0.409\pm0.015$ &... \\
$\chi^2_\nu(\nu)$ & ... &$1.29(100)$&$0.96(360)$&... \\
\hline
\multicolumn{4}{l}{Model: optxagnf (maximally-rotating,  pure thermal disk)}  \\
$N_\mathrm{H,i}$ (10$^{20}$ cm$^{-2}$)\textsuperscript{a} & ... &\multicolumn{2}{c}{$1.5\pm1.0$}&... \\
$M_\mathrm{BH}$ (\msun)& ... & \multicolumn{2}{c}{ $ 1.2\pm0.1\times10^5$} &... \\
$L_\mathrm{bol}/L_\mathrm{Edd}$ & ... & $0.406\pm0.014$ &$0.167\pm0.006$ & ... \\
$\chi^2_\nu(\nu)$ & ... &$1.29(100)$&$0.96(360)$&... \\
\hline
\multicolumn{4}{l}{Model: nsatmos ($M_\mathrm{NS}=1.4$ \msun, $R_\mathrm{NS}=10$ km)}\\
$N_\mathrm{H,i}$ (10$^{20}$ cm$^{-2}$)\textsuperscript{a} & \multicolumn{4}{c}{$3.3\pm0.8$}\\
$kT_\mathrm{eff}^\infty$ (eV) & $ 74.6\pm0.5$ & $ 70.0\pm0.3$ & $ 55.7\pm0.5$ & $ 37.5\pm0.9$\\
$D_\mathrm{nsatmos}$ (kpc)\textsuperscript{a} &  \multicolumn{4}{c}{$2.2\pm0.2$} \\
$\chi^2_\nu(\nu)$ & 1.10(327) &1.34(102) &1.02(361) &3.18(8) \\
$L_\mathrm{bol}$ ($10^{32}$ erg s$^{-1}$) &$6.78\pm0.18$ & $5.26\pm0.10$ &$2.11\pm0.08$ &$0.43\pm0.04$\\
\hline
\multicolumn{4}{l}{Model: nsatmos ($M_\mathrm{NS}=1.4$ \msun, $R_\mathrm{NS}=13$ km)}\\
$N_\mathrm{H,i}$ (10$^{20}$ cm$^{-2}$)\textsuperscript{a} & \multicolumn{4}{c}{$3.1\pm0.8$}\\
$kT_\mathrm{eff}^\infty$ (eV) & $ 76.3\pm0.5$ & $ 71.5\pm0.4$ & $ 57.0\pm0.5$ & $ 38.3\pm0.9$\\
$D_\mathrm{nsatmos}$ (kpc)\textsuperscript{a} &  \multicolumn{4}{c}{$2.7\pm0.2$} \\
$\chi^2_\nu(\nu)$ & 1.11(327) &1.34(102) &1.03(361) &3.14(8) \\
$L_\mathrm{bol}$ ($10^{32}$ erg s$^{-1}$) &$10.80\pm0.29$ & $8.36\pm0.17$ &$3.36\pm0.12$ &$0.69\pm0.06$\\
\hline
\multicolumn{4}{l}{Model: nsatmos ($M_\mathrm{NS}=1.4$ \msun, $R_\mathrm{NS}=16$ km)}\\
$N_\mathrm{H,i}$ (10$^{20}$ cm$^{-2}$)\textsuperscript{a} & \multicolumn{4}{c}{$2.8\pm0.8$}\\
$kT_\mathrm{eff}^\infty$ (eV) & $ 78.5\pm0.5$ & $ 73.6\pm0.4$ & $ 58.6\pm0.5$ & $ 39.3\pm0.9$\\
$D_\mathrm{nsatmos}$ (kpc)\textsuperscript{a} &  \multicolumn{4}{c}{$3.5\pm0.3$} \\
$\chi^2_\nu(\nu)$ & 1.11(327) &1.34(102) &1.03(361) &3.13(8) \\
$L_\mathrm{bol}$ ($10^{32}$ erg s$^{-1}$) &$16.90\pm0.45$ & $13.08\pm0.25$ &$5.22\pm0.19$ &$1.06\pm0.10$\\
\hline
\hline
\end{tabular}
\end{table*}

\clearpage

\begin{figure*}[!tb]
\centering
\includegraphics[width=5.4in]{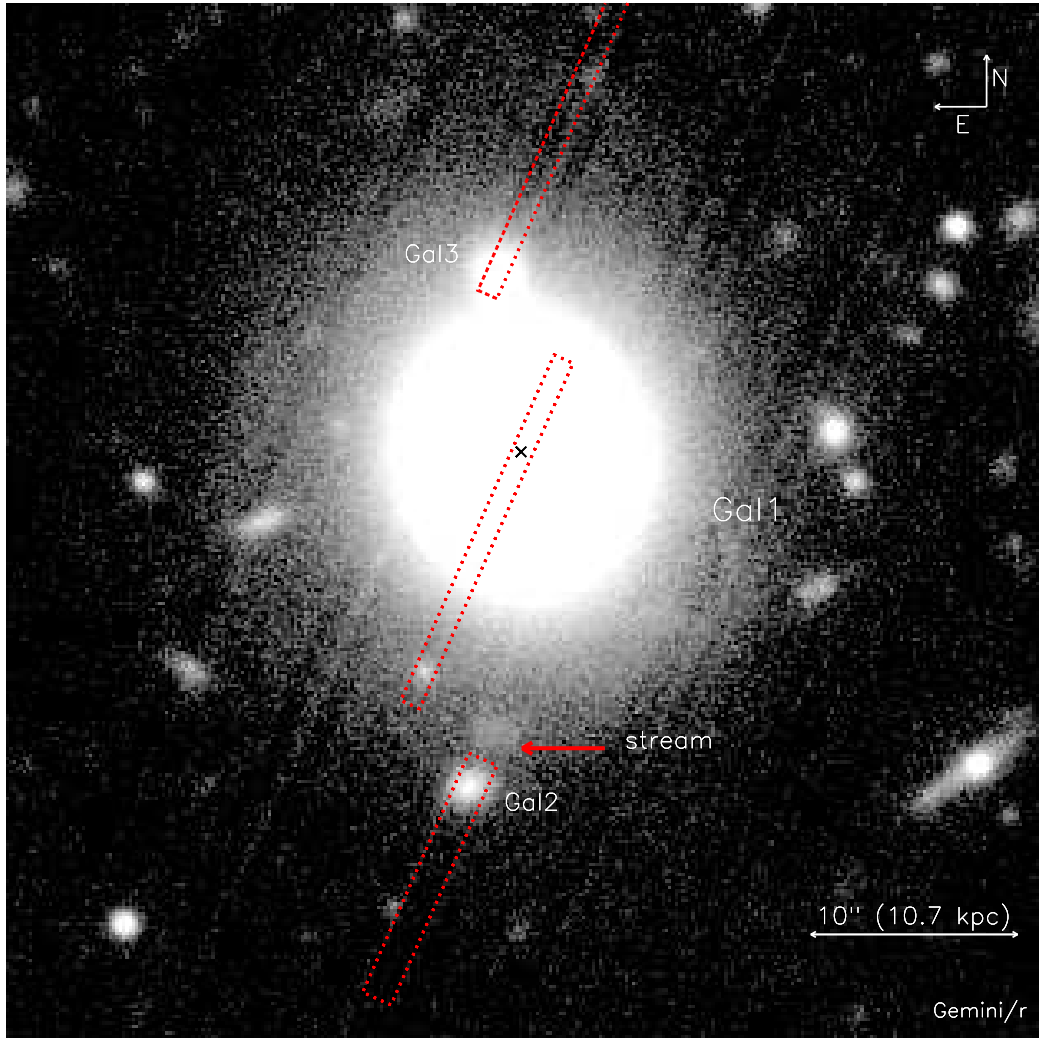}
\caption{\textbf{The Gemini image used to make the slits for a
    GMOS observation around the field of J2150$-$0551.} Three slits are shown here, with one going through the nucleus of Gal1 and the counterpart to J2150$-$0551, one for Gal2, and one for a possible background galaxy Gal3 (marked in the plot, not studied in this work). \label{fig:geminiimg}}
\end{figure*}

\clearpage

\begin{figure*}[!tb]
\centering
\includegraphics[width=3.4in]{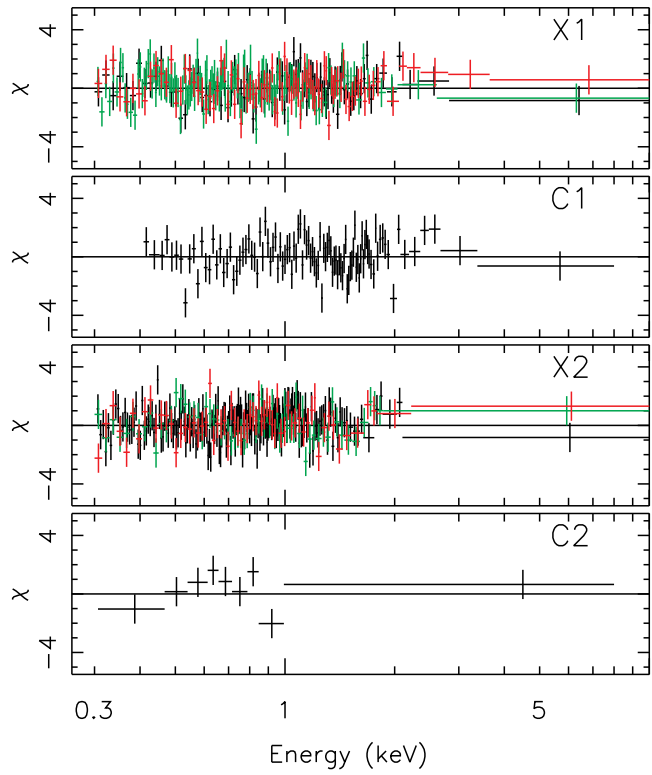}
\caption{\textbf{The residuals of the absorbed \textit{diskbb} fits shown in Figure~\ref{fig:srcspec_disk_delchi}, in units of standard deviations with error bars of size one.} The pn, MOS1, and MOS2 data in X1 and X2 are shown in black, red, and green, respectively. \label{fig:srcspec_disk_delchi}}
\end{figure*}
\clearpage

\begin{figure*}
\centering
\includegraphics[width=3.4in]{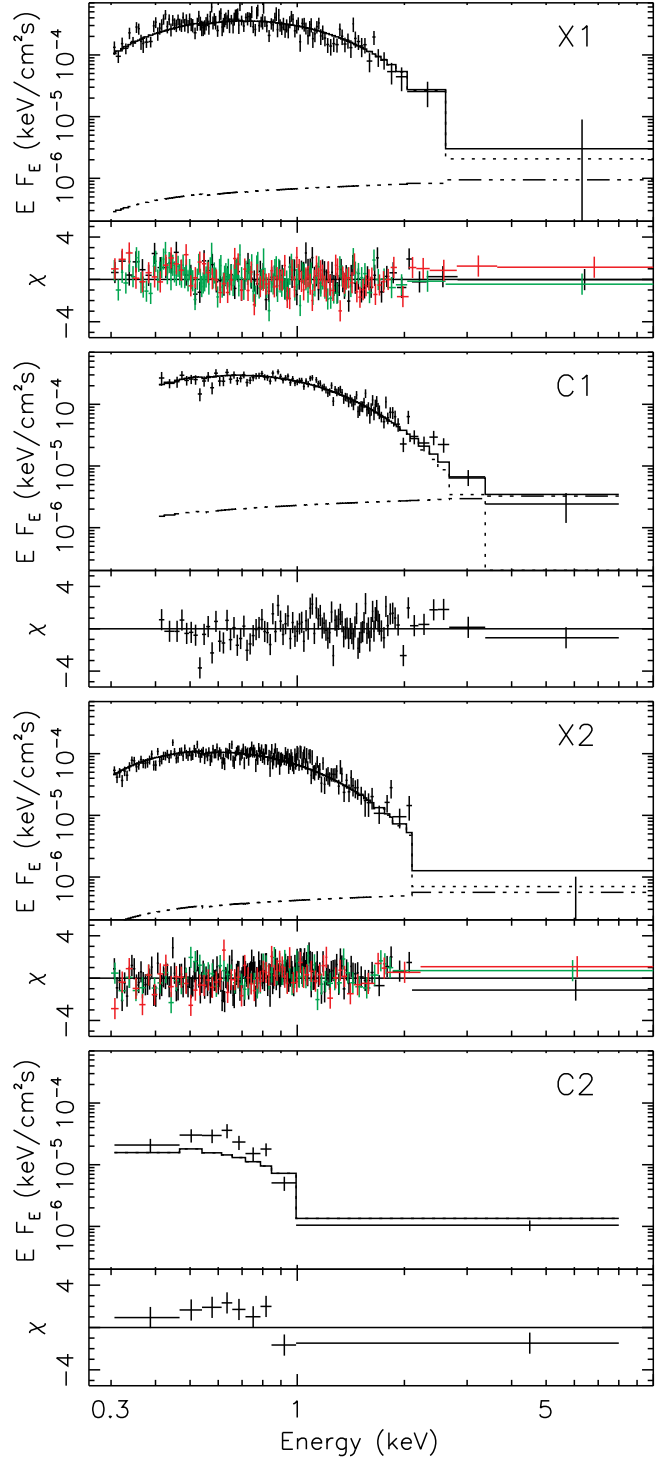}
\caption{\textbf{The unfolded spectra (black data points with 1$\sigma$ error bars) and fit residuals (in units of standard deviations with error bars of size one) of {\it XMM-Newton}
  and {\it Chandra} observations, similar to the \textit{diskbb} fits shown in
  Figure~\ref{fig:lumlcsp} and Supplementary
  Figure~\ref{fig:srcspec_disk_delchi} but for an absorbed \textit{nsatmos}
  model.} Note that the fits to the X1, C1, and X2 spectra also include
a PL of $\Gamma_\mathrm{PL}=1.8$ to account for possible contamination
from the faint nuclear source of Gal1 (Source 1). All error bars represent 1$\sigma$ uncertainties. \label{fig:srcspec_nsatmos_withdelchi}} 
\end{figure*}

\clearpage

\begin{figure*}[!tb]
  \centering
\includegraphics[width=5.4in]{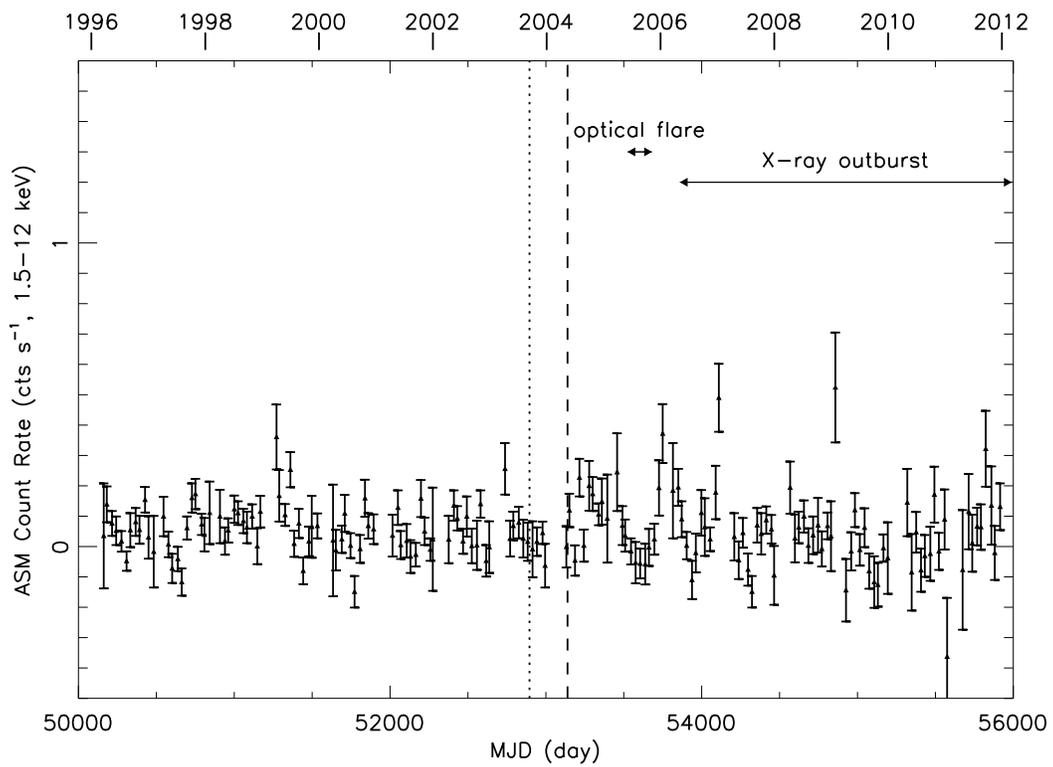}
\caption{\textbf{The ASM monthly light curve of J2150$-$0551,
    indicating no significant emission at any time of the mission.} Error bars represent 1$\sigma$ uncertainties.
  The dotted vertical line marks the time of the \emph{HST}/ACS F775W
  observation (the optical flare should have started later than this),
  and the dashed vertical line marks the time of the \emph{XMM-Newton}
  slew observation on 14 May 2004 (the accretion outburst of
  J2150$-$0551, if it is an NS LMXB, should have started later than this). Time ranges of the optical flare detected by the CFHT and the X-ray outburst detected by \emph{XMM-Newton} and \emph{Chandra} are also marked. The X-ray outburst was not detected by the ASM because its peak flux was three orders of magnitude below the ASM sensitivity limit. \label{fig:asmlightcurve}}
\end{figure*}
\clearpage

\begin{figure*}[!tb]
\centering
\includegraphics[width=3.4in]{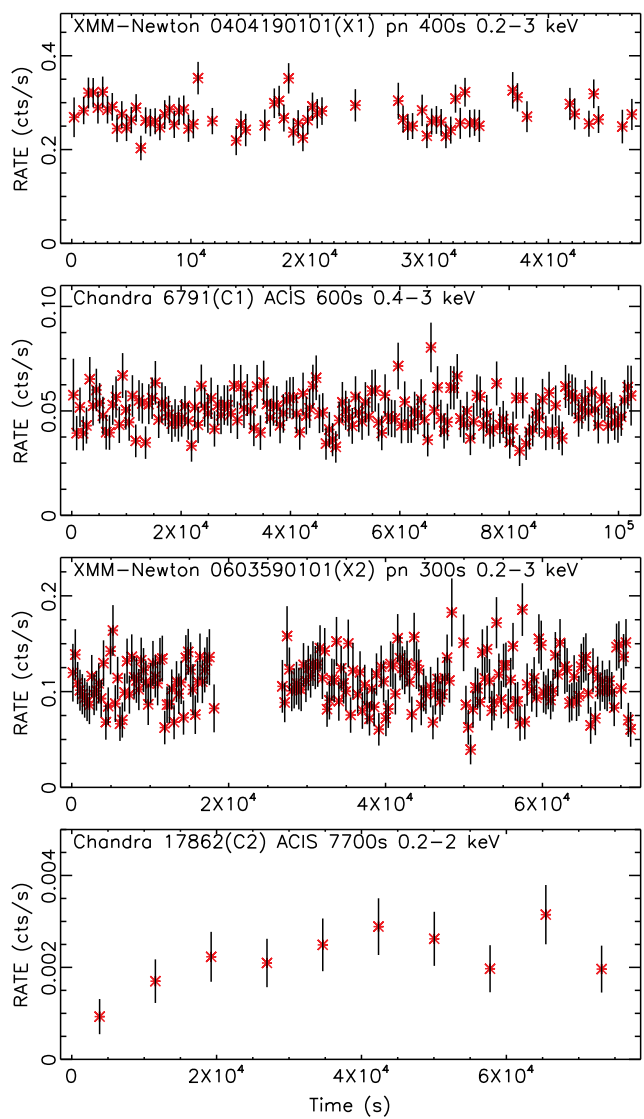}
\caption{\textbf{The background subtracted light curves of J2150$-$0551 from X1,
  C1, X2, and C2.} Error bars represent 1$\sigma$ uncertainties. The instrument, light curve bin size, and energy band
  used for each light curve are noted in the plot. The source showed no
  significant short-term variability in any
  observation. \label{fig:srclc}}
\end{figure*}
\clearpage

\begin{figure*}[!tb]
\centering
\includegraphics[width=6.8in]{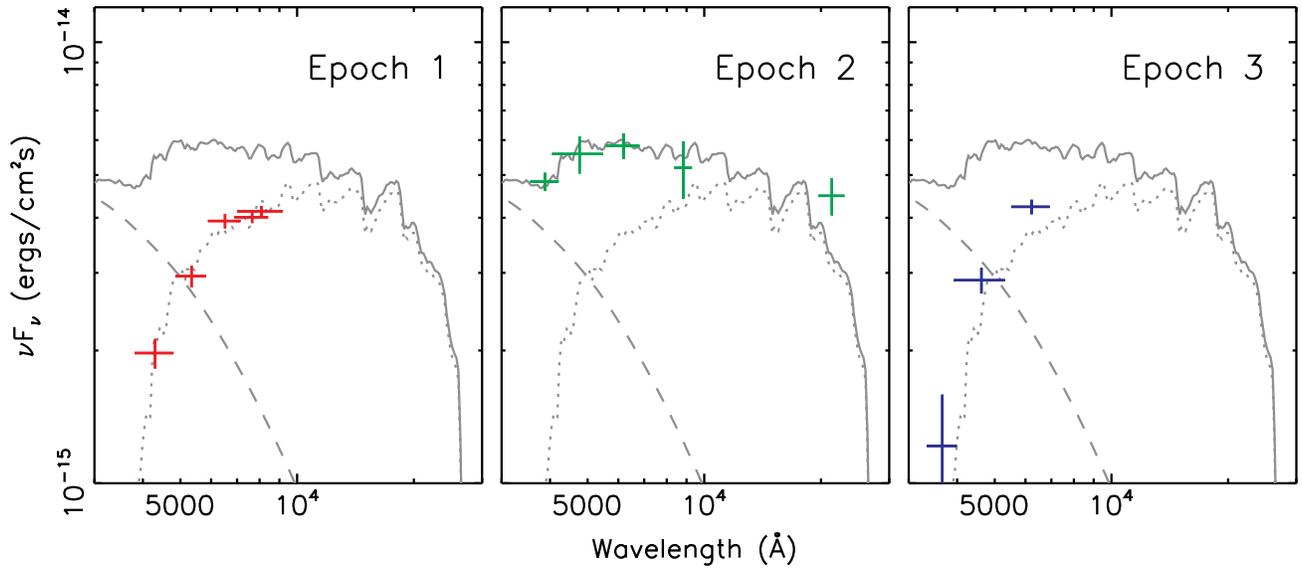}
\caption{\textbf{The optical/IR photometry (and 1$\sigma$ error bars) of J2150$-$0551 in various epochs, indicating the presence of an optical flare in 2005.} Left panel: the
  CFHT/CFH12K (2000--2001) and \textit{HST}/ACS F775W (2003) photometry. Middle panel: the CFHT/MegaCam optical (2005) and WIRCam Ks (2009) photometry (green
  data). Right panel: the SOAR, Subaru, and Gemini
  photometry in 2014--2016. The gray solid line in all panels shows the fit to the broad-band spectrum combining X1, the CFHT/MegaCam and WIRCam photometry, with an irradiated disk (gray dashed line) plus a stellar population emission component (gray dotted line, inferred from the CFHT/CFH12K and \textit{HST}/ACS photometry in 2000--2003). \label{fig:srcspec_diskirspp}}
\end{figure*}

\clearpage

\begin{figure*}[!tb]
\centering
\includegraphics[width=5.4in]{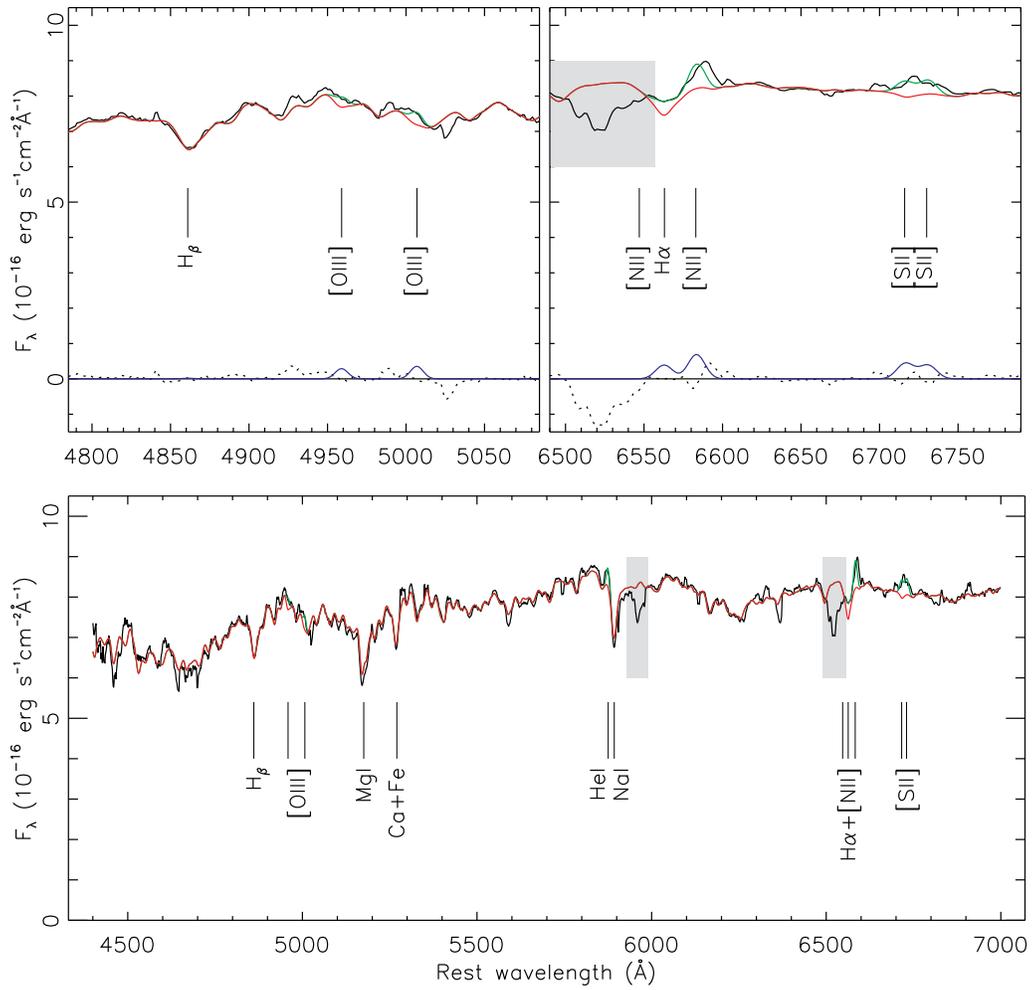}
\caption{\textbf{The pPXF fit to the Gemini spectrum around the nuclear region of Gal1, with the pPXF fit shown as a solid green line and the star component shown as a red line.} The upper two panels zoom into the H$\beta$-[\ion{O}{III}] complex and the H$\alpha$-[\ion{N}{II}] region, including the fit residuals (dotted lines) and the Gaussian gas lines (blue solid lines). The gray areas mark the regions where the spectrum was seriously affected by the atmospheric OH absorption and CCD gap. \label{fig:geminispfit_Gal1}}
\end{figure*}
\clearpage

\begin{figure*}[!tb]
\centering
\includegraphics[width=5.4in]{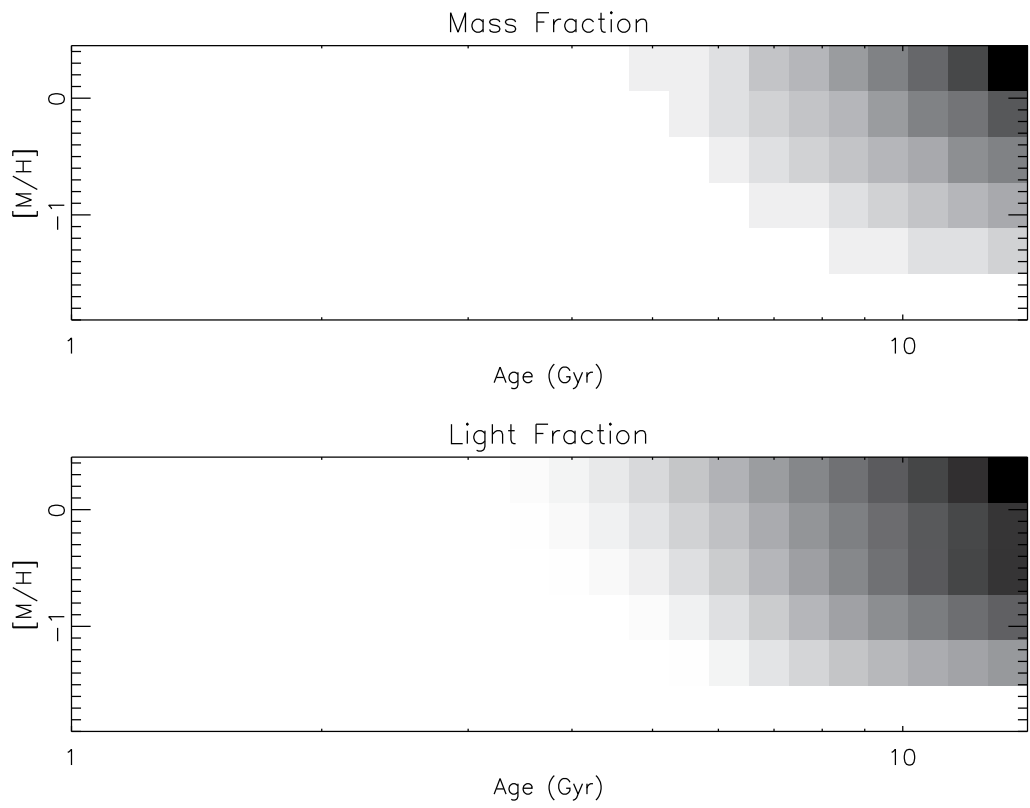}
\caption{\textbf{Relative mass and light fractions of stellar
  populations in the nuclear region of Gal1 with respect to
  metallicity and age.} Darker shading indicates a larger mass
  fraction in the best-fitting model. The light was integrated over 4400--7000 \AA. \label{fig:masslightdis_Gal1}}
\end{figure*}
\clearpage

\begin{figure*}[!tb]
\centering
\includegraphics[width=5.4in, trim={0 0 0 3.2in},clip]{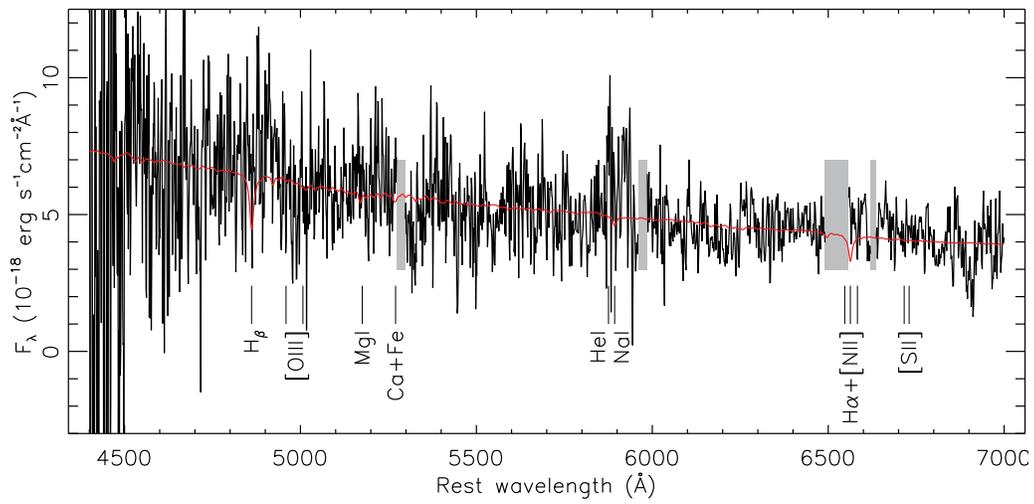}
\caption{\textbf{The pPXF fit to the Gemini spectrum of Gal2, assuming it to be at the same redshift as Gal1.} The spectrum is somewhat noisy, and we observed no significant emission lines or absorption features. The gray areas mark some portions of the spectra seriously affected by the atmospheric OH absorption and CCD gaps. \label{fig:geminispfit_G2}}
\end{figure*}
\clearpage

\begin{figure*}[!tb]
\centering
\includegraphics[width=5.4in]{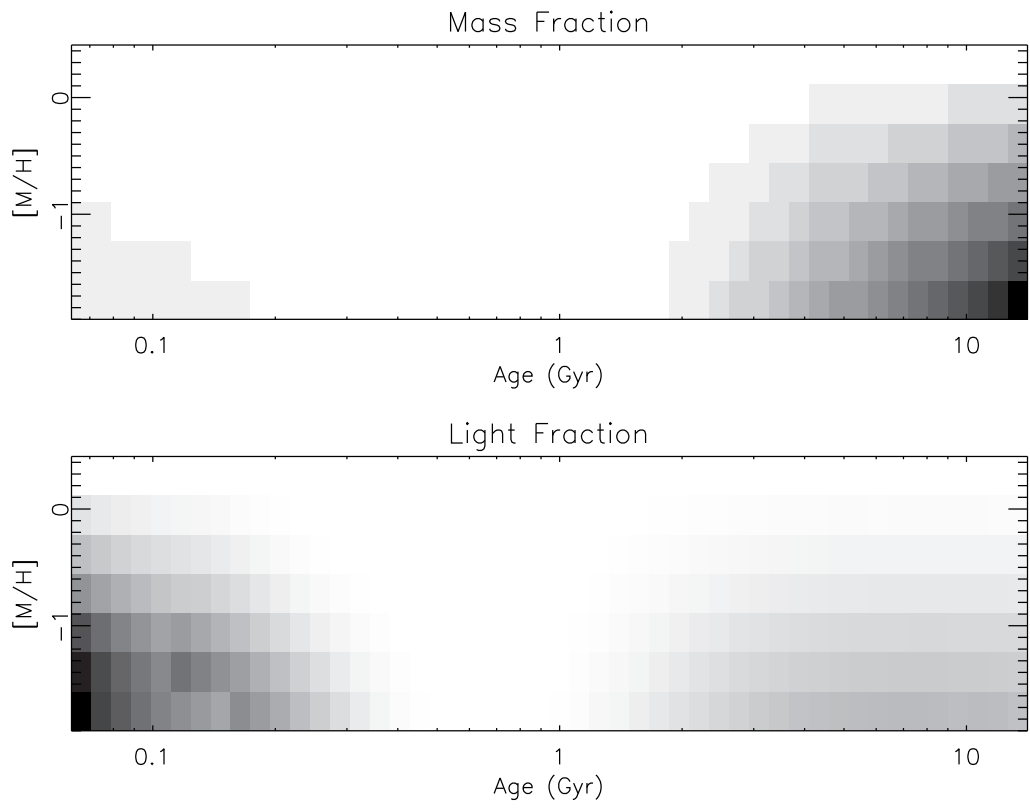}
\caption{\textbf{Relative mass and light fractions of stellar
  populations in Gal2 with respect to
  metallicity and age, suggesting the presence of young stellar populations of
    $\lesssim$0.3 Gyr.} \label{fig:masslightdis_G2}}
\end{figure*}
\clearpage

\begin{figure*}[!tb]
\centering
\includegraphics[width=5.4in]{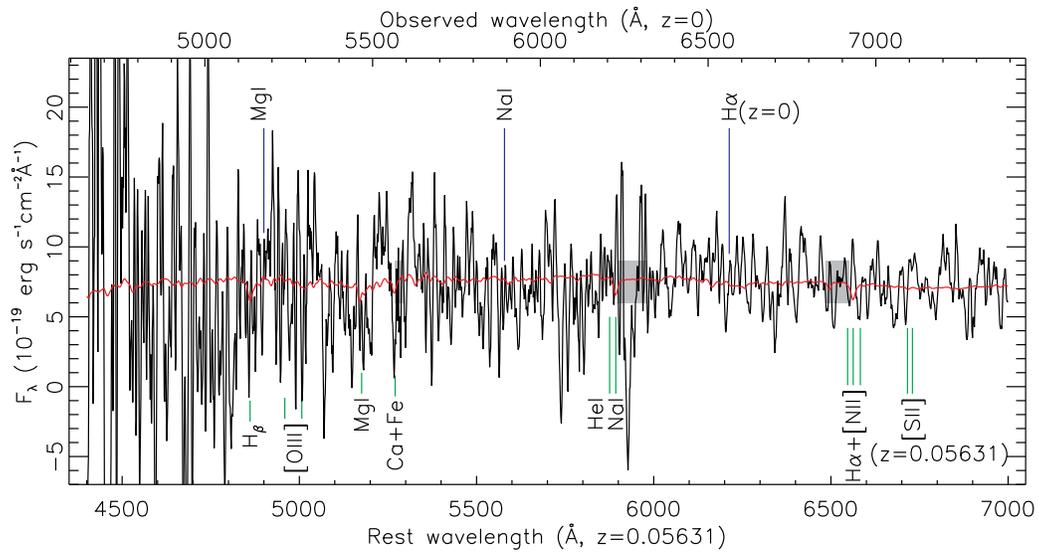}
\caption{\textbf{The pPXF fit to the Gemini spectrum of the
    counterpart to J2150$-$0551, showing no significant emission lines
    or absorption features.} We assume the spectrum to be blueshifted
  by 300 km s$^{-1}$ relative to Gal1 in the pPXF fit. The expected typical AGN emission lines and stellar absorption features are also marked below the spectrum. We also mark the expected typical emission/absorption lines in the case that J2150$-$0551 is a Galactic object with zero redshift. The spectrum is noisy and has been smoothed with a box function of width 8.6\AA\ (the FWHM resolution of the spectrum) for clarity. The gray areas mark some portions of the spectrum seriously affected by the atmospheric OH absorption and CCD gaps. \label{fig:geminispfit_J2150$-$0551}}
\end{figure*}
\clearpage

\begin{figure*}
\centering
\includegraphics{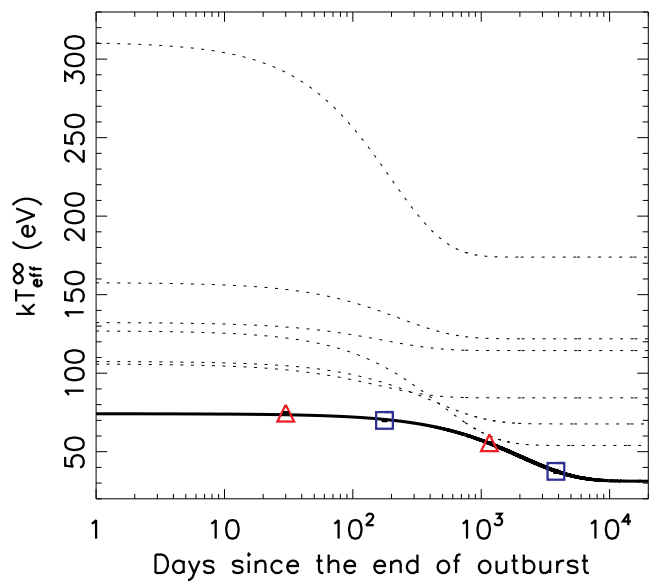}
\caption{\textbf{Evolution of the effective temperature based on the \textit{nsatmos} fits to the X-ray spectra of J2150$-$0551, assuming an NS of mass 1.4 \msun\ and radius 10 km.} The error bars, smaller than the symbol size, are at the 90\% confidence level. The dotted lines plot the fits of the cooling curves from six cooling NSs (MAXI J0556$-$332, XTE J1701$-$462, EXO 0748$-$676, MXB 1659$-$29, KS 1731$-$260, IGR J17480$-$2446, from the top to the bottom at the time of day 1)\cite{hofrwi2014} with an exponential decay to a constant ($T_\mathrm{eff}^\infty(t) = T_1 \exp^{-t/\tau} + T_0$). For J2150$-$0551, the ending date of the hypothetical accretion outburst is unknown, and we assumed it to be one month before X1 when we fitted the cooling curve (solid line). The fit inferred a decaying timescale of 2008 days, which is independent of the ending date of the accretion outburst assumed.  \label{fig:coolingcurve} }
\end{figure*}

\end{document}